\documentclass[preprint]{aastex}
\usepackage{color,natbib}
\usepackage{lscape}

\newcommand{\Spitzer}{{\sl Spitzer}}

\newcommand{\HST}{{\sl HST}}

\newcommand{\Msun}{\mbox{$M_{\sun}$}}

\newcommand{\Lsun}{\mbox{$L_{\sun}$}}

\newcommand{\Mjup}{\mbox{$M_{\rm Jup}$}}

\newcommand{\etal}{et al.}
\newcommand{\eg}{e.g.}
\newcommand{\ie}{i.e.}

\newcommand{\meth}{{\hbox{CH$_4$}}}   

\newcommand{\Ks}{\mbox{$K_S$}}
\newcommand{\degs}{\mbox{$^{\circ}$}}
\newcommand{\Lbol}{\mbox{$L_{\rm bol}$}}

\newcommand{\Teff}{\mbox{$T_{\rm eff}$}}
\newcommand{\logg}{\mbox{$\log(g)$}}

\newcommand{\Rhk}{\mbox{$R^{\prime}_{HK}$}}

\newcommand{\eInd}{\hbox{$\epsilon$~Ind}}
\newcommand{\eIndBab}{\hbox{$\epsilon$~Ind~Bab}}
\newcommand{\twomassbinAB}{\hbox{2MASS~J1209$-$1004AB}}
\newcommand{\twomassbin}{\hbox{2MASS~J1209$-$1004}}

\slugcomment{ApJ, in press}
\shorttitle{Highly Unequal-Mass Binary T Dwarfs}
\shortauthors{Liu \etal}

\begin{document}

\title{Discovery of a Highly Unequal-Mass Binary T~Dwarf with Keck Laser
  Guide Star Adaptive Optics: A Coevality Test of Substellar Theoretical
  Models and Effective Temperatures\altaffilmark{1}}


\author{Michael C. Liu,\altaffilmark{2} 
  Trent J. Dupuy,\altaffilmark{2} and
  S. K. Leggett\altaffilmark{3}}

\altaffiltext{1}{Most of the data presented herein were obtained at the
  W.M. Keck Observatory, which is operated as a scientific partnership
  among the California Institute of Technology, the University of
  California, and the National Aeronautics and Space Administration. The
  Observatory was made possible by the generous financial support of the
  W.M. Keck Foundation.}
\altaffiltext{2}{Institute for Astronomy, University of Hawai`i, 2680
  Woodlawn Drive, Honolulu, HI 96822; mliu@ifa.hawaii.edu}
\altaffiltext{3}{Gemini Observatory, 670 North A'ohoku Place, Hilo, HI
  96720}

\begin{abstract} 
  \noindent Highly unequal-mass ratio binaries are rare
    among field brown dwarfs, with the mass ratio distribution of the
    known census described by $q^{(4.9\pm0.7)}$. However, such systems
    enable a unique test of the joint accuracy of evolutionary and
    atmospheric models, under the constraint of coevality for the
    individual components (the ``isochrone test''). We carry out this
    test using two of the most extreme field substellar binaries
    currently known, the T1+T6 \eIndBab\ binary and a newly discovered
    0.14\arcsec\ T2.0+T7.5 binary, 2MASS~J12095613$-$1004008AB,
    identified with Keck laser guide star adaptive optics. The latter is
    the most extreme tight binary resolved to date ($q\approx0.5$).
    Based on the locations of the binary components on the H-R diagram,
    current models successfully indicate that these two systems are
    coeval, with internal age differences of $\log(age)=-0.8\pm1.3\
    (-1.0_{-1.3}^{+1.2}$)~dex and $0.5^{+0.4}_{-0.3}\
    (0.3^{+0.3}_{-0.4})$~dex for \twomassbinAB\ and \eIndBab,
    respectively, as inferred from the Lyon (Tucson) models.
  However, the total mass of \eIndBab\ derived from the H-R diagram
  ($\approx$80~\Mjup\ using the Lyon models) is strongly discrepant with
  the reported dynamical mass.
  This problem, which is independent of the assumed age of the \eIndBab\
  system, can be explained by a $\approx$50--100~K
    systematic error in the model atmosphere fitting, indicating
    slightly warmer temperatures for both components; bringing the mass
    determinations from the H-R diagram and the visual orbit into
  consistency leads to an inferred age of $\approx$6~Gyr for \eInd~Bab,
  older than previously assumed. Overall, the two T~dwarf binaries
  studied here, along with recent results from T~dwarfs
    in age and mass benchmark systems, yield evidence for small
    ($\approx$100~K) errors in the evolutionary models and/or model
    atmospheres, but not significantly larger.
  Future parallax, resolved spectroscopy,
 and dynamical mass measurements for \twomassbinAB\ will enable a more
  stringent application of the isochrone test.  
  Finally, the binary nature of this object reduces its utility as
  the primary T3 near-IR spectral typing standard; we suggest
  SDSS~J1206+2813 as a replacement. \end{abstract}



\section{Introduction}

Wide-field surveys have now identified nearly a thousand ultracool
dwarfs (spectral types later than M6) in the solar neighborhood,
spanning temperatures and (inferred) masses that range from the
stellar/substellar boundary defined by the hydrogen-burning minimum
mass ($\approx$75~\Mjup), down to the proposed substellar/planetary
boundary demarcated by the deuterium-burning limit ($\approx$13~\Mjup).
This census has enabled extensive characterization of the
spectrophotometric properties of these low-luminosity, low-temperature
objects. However, studies of the substellar field population are
inevitably hindered by the unknown properties of individual objects,
making it difficult to disentangle the effects of varying ages, masses,
and compositions on the underlying physics.
In principle, a sample of star clusters with complete membership down to
very low masses and encompassing a range of ages would be ideal.
However, such a sensitive census only exists for a handful of young
($\sim$few~Myr) clusters \citep[e.g.][]{2007MNRAS.380..712L,
  2007prpl.conf..443L}, where the lowest mass members are easiest to
detect. With increasing age, dynamical evolution leads to depletion of
the lowest mass members such that older clusters may be largely devoid
of brown dwarfs \citep{2002MNRAS.333..547A, 2008A&A...481..661B}.

One avenue for circumventing these limitations is through the study of
binary brown dwarfs, as binaries constitute systems of common (albeit
unknown) age and metallicity. About 15\% of field brown dwarfs are tight
binary systems, resolved largely by {\sl Hubble Space Telescope} or
ground-based adaptive optics (AO) imaging and composed mostly of nearly
equal-luminosity (and thus equal-mass) components
\citep[e.g.][]{2006astro.ph..2122B}. These ``mini-clusters'' have proven
to be fertile ground for understanding the physical processes that
govern the emergent spectra of brown dwarfs
\citep[e.g.][]{2005astro.ph..8082L, 2006astro.ph..5037L,
  2006ApJS..166..585B} and for testing theoretical models with dynamical
mass determinations \citep[e.g.][]{2004astro.ph..7334O, gl802b-ireland,
  liu08-2m1534orbit, 2009ApJ...699..168D, konopacky10-binaries,
  2010arXiv1007.4197D}.

As part of our ongoing program to study the multiplicity and physical
properties of substellar binaries using laser guide star adaptive optics
(LGS AO), we present here the discovery of the binarity of
2MASS~J12095613$-$1004008, hereinafter \twomassbin. This object was
identified in 2MASS data by \citet{2004AJ....127.2856B} and spectrally
typed by them and by \cite{chiu05} as a T3 dwarf based on
integrated-light near-IR spectroscopy. \citet{2005astro.ph.10090B}
define \twomassbin\ as the primary T3 spectral type standard for their
near-IR classification scheme for T~dwarfs. This object has not
previously been targeted with high angular resolution imaging.
What distinguishes \twomassbinAB\ from most previously known
substellar field binaries is the large IR brightness difference
between its two components, indicating an atypical mass ratio compared
to the plethora of nearly equal-mass binaries. As we describe below,
\twomassbinAB\ provides a new opportunity to test theoretical models
of brown dwarfs --- by requiring the model-derived ages of the two
components to be consistent with coevality of the system, an approach
which we refer to here as the ``isochrone test.''

\section{Observations}

We imaged \twomassbin\ on 22~April~2007 and 16~January~2008~UT using the
sodium LGS AO system of the 10-meter Keck II Telescope on Mauna Kea,
Hawaii \citep{2006PASP..118..297W, 2006PASP..118..310V}. We used the
facility IR camera NIRC2 with its narrow field-of-view camera, which
produces a $10.2\arcsec \times 10.2\arcsec$ field of view. Conditions
were photometric for both runs. The LGS provided the wavefront reference
source for AO correction, with the tip-tilt motion measured
contemporaneously from the $R=14.7$~mag field star
USNO-B1.0~0799-0230529 \citep{2003AJ....125..984M} located 68\arcsec\
away from \twomassbin.
The LGS brightness, as measured by the flux incident on the AO wavefront
sensor, was equivalent to a $V\approx 9.6-10.0$~mag star. 
In April~2007, we obtained images with the MKO $J$ (1.25~\micron), $H$
(1.64~\micron), and $K$ (2.20~\micron) filters. In January 2008, we
obtained images with the $CH_4s$ filter, which has a central wavelength
of 1.592~\micron\ and a width of 0.126~\micron.

The images were analyzed in the same fashion as our previous LGS papers
\citep{liu08-2m1534orbit, 2006astro.ph..5037L}. The raw images were
reduced using standard methods of flat-fielding and sky-subtraction. The
binary's flux ratios and relative astrometry were derived by fitting an
analytic model of the point spread function (PSF) as the sum of three
elliptical gaussians. The measurements were made on the individual
images, with outlier images of much poorer quality excluded. The
averages of the fitting results were adopted as the final measurements
and the standard deviations as the errors. The final flux ratios were
used to correct the Strehl ratio measurements of the images for the
contamination from the light of the secondary component. We did not
correct the relative astrometry for instrumental optical distortion, as
the size of the effect as predicted from a distortion solution by B.
Cameron (priv. comm.) is insignificant compared to our measurements
uncertainties.

In order to validate our measurements, we created myriad artificial
binary stars from LGS images of single stars with comparable Strehl and
FWHM as the science data. For each filter, our fitting code was applied
to artificial binaries with similar separations and flux ratios as
\twomassbinAB\ over a range of PAs. The simulations showed that the
random errors are reasonable and any systematic offsets are small.
In cases where the measurement errors from the artificial binaries were
larger than those from the \twomassbinAB\ measurements, we
conservatively adopted the larger errors.

\begin{figure}
\vskip -4in
\centerline{\includegraphics[height=6.5in,angle=90]{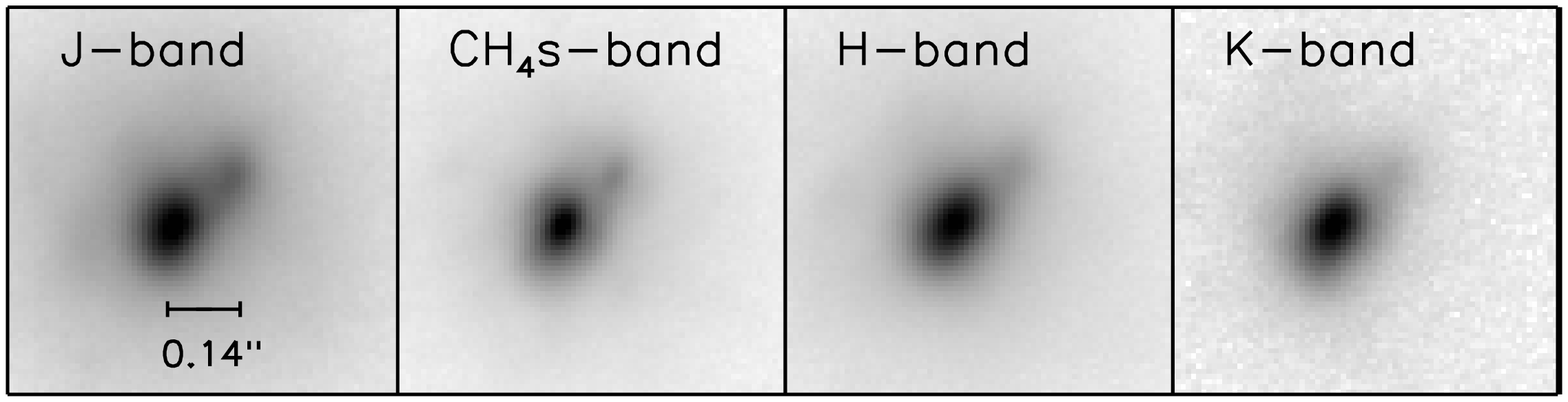}}
\vskip -3in
\centerline{\includegraphics[height=6.5in,angle=90]{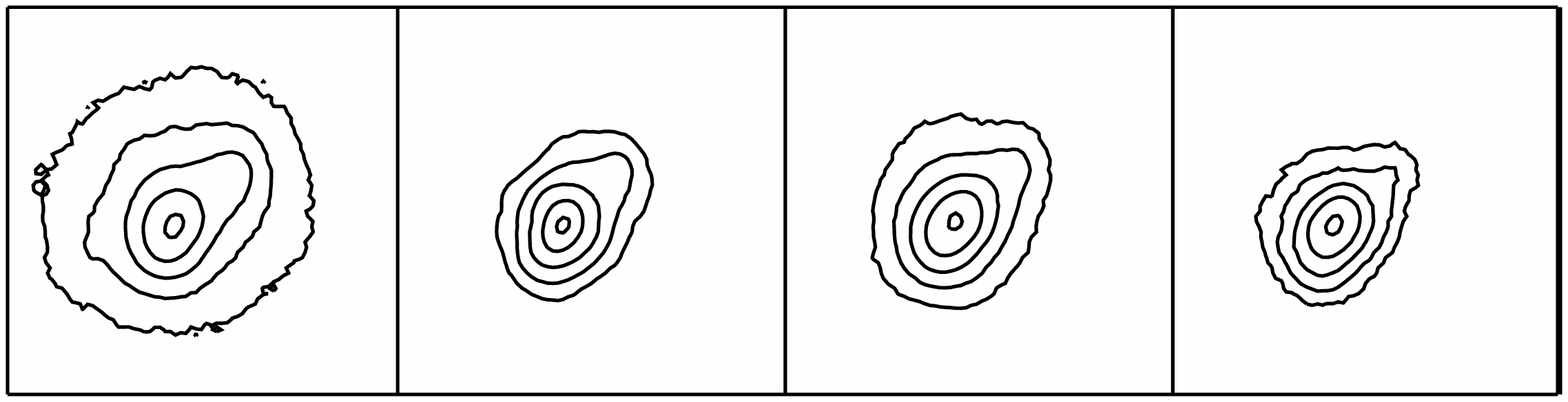}}
\caption{\normalsize Keck LGS AO imaging of \twomassbin{AB}.  North is
  up and east is left.  Each image is 0.75\arcsec\ on a side. Contours
  are drawn from 90\%, 45\%, 22.5\%, 11.2\%, and 5.6\% of the peak
  value in each bandpass. \label{fig:images}}
\end{figure}

We adopted a pixel scale of $9.963 \pm 0.005$~mas/pixel and an
orientation for the detector's $+y$~axis of $+0.13 \pm 0.07$\degs\ for
NIRC2 as determined by \citet{2008ApJ...689.1044G}.
We computed the expected shift in the relative astrometry of the two
components due to differential chromatic refraction (DCR) in the same
manner as in our previous work \citep[e.g.][]{2009ApJ...699..168D}. The
DCR effects are small ($\approx$1--3~mas) compared to the measurement
errors, and therefore we did not account for them.
Table~\ref{table:keck} presents our final Keck LGS measurements, and
Figure~\ref{fig:images} shows our Keck LGS images.


\section{Results for \twomassbinAB}

\subsection{Proof of Companionship}

The near-IR broadband colors of \twomassbin{B} are extremely blue,
which provides strong circumstantial evidence that it is a physically
associated late-T dwarf companion and not a background object.
\citet{2004AJ....127.2856B} measure a proper motion for \twomassbin\
of $0.46\pm0.10$\arcsec/yr at a PA of $140\pm8\degs$. Considering this
proper motion, examination of the Digitized Sky Survey images shows no
plausible optical counterpart if component~B were a background object.
Our two epochs of Keck LGS imaging separated by 9~months are
consistent with no change in the relative position of the two
components. If the companion were a background object, in Jan 2008 the
system should have had a separation of $0.48\pm0.07$\arcsec\ at a PA
of $317\pm6$\degs, which is $4.8\sigma$ discrepant with the observed
separation, neglecting the effect of parallax.
Finally, our LGS photometry indicates the presence of $H$-band methane
absorption in component~B (\S~\ref{sec:phot}), inconsistent with the
background star hypothesis. Thus both the astrometry and photometry
independently indicate that the system is a true physical binary.

\subsection{Resolved Photometry\label{sec:phot}}

\begin{figure}
\vskip -1in
\hskip 0.2in
\centerline{\includegraphics[width=6in,angle=90]{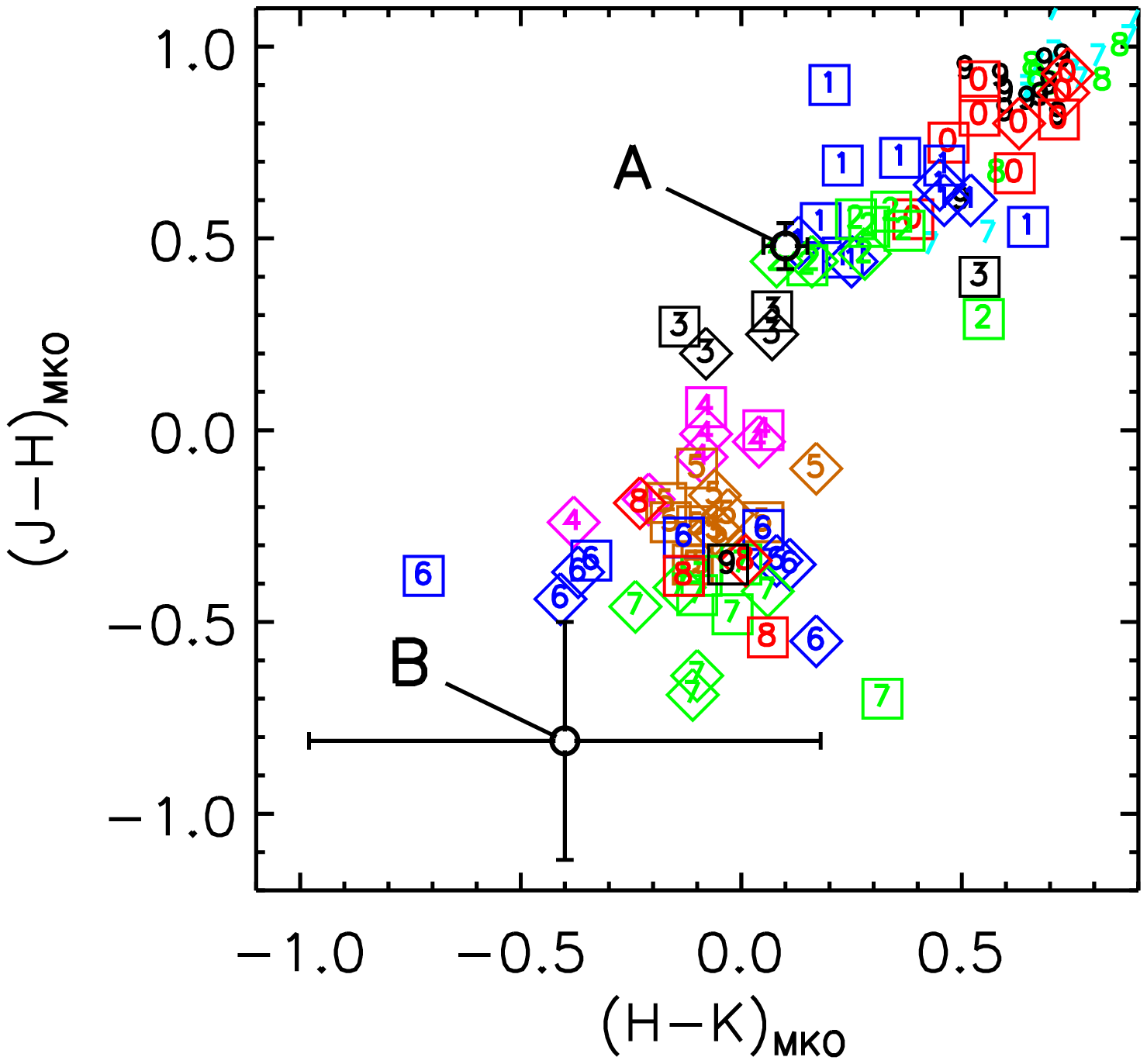}}
\vskip -2ex
\caption{\normalsize Resolved near-IR colors of the components of
  \twomassbinAB\ compared with single late-L~and T~dwarfs from
  \citet{2004AJ....127.3553K} and \citet{chiu05}; late-T dwarfs
  ($\ge$T6) from \citet{2007MNRAS.381.1400W},
  \citet{2008MNRAS.385L..53C}, \citet{2008MNRAS.390..304P},
  \citet{2008A&A...482..961D}, and \citet{2008MNRAS.391..320B}; and
  individual components of resolved binaries from
  \citet{2004A&A...413.1029M}, \citet{2005astro.ph.10580B},
  \citet{2006ApJS..166..585B}, \citet{2005astro.ph..8082L}, and
  \citet{2006astro.ph..5037L}.  (For the T8.5 dwarf CFBDS~0059$-$0114,
  the $K$-band measurement is synthesized from the near-IR spectrum
  and \Ks-band photometry of \citealp{2008A&A...482..961D}.)  The
  photometry errors are comparable to or smaller than the size of the
  plotting symbols for most of the objects.  The numbers indicate the
  near-IR spectral subclass of the objects, with half subclasses being
  rounded down (\eg, T3.5 is labeled as ``3''), and objects of the
  same subclass plotted in the same color.  The late-L~dwarfs
  (classified on the \citealp{geb01} scheme) are plotted as bare
  numbers.  The T~dwarfs (on the \citealp{2005astro.ph.10090B} scheme
  with extension by \citealp{2008MNRAS.391..320B}) are plotted as
  circumscribed numbers, with squares for integer subclasses (\eg, T3)
  and diamonds for half subclasses (\eg, T3.5).
  \label{fig:colorcolor}}
\end{figure}

We use our measured flux ratios and the published $JHK$ photometry
from \cite{chiu05} to derive resolved IR colors and magnitudes for
\twomassbinAB\ on the MKO system.  For the $CH_4s$ data, we synthesize
integrated-light photometry of $15.05\pm0.03$~mag from the near-IR
spectrum of \citet{2004AJ....127.2856B}, flux-calibrated to the
broadband MKO photometry of $H = 15.24\pm0.03$~mag.  We assume that
the components of \twomassbinAB\ are themselves single, not unresolved
binaries.  Table~\ref{table:resolved} presents the resolved
photometry.
Comparing to known ultracool dwarfs, Figure~\ref{fig:colorcolor} shows
that component~A has IR colors most typical of T2--T3 dwarfs.  The
uncertainties are larger for component~B, but its near-IR colors
appear to be comparable or even bluer than the very latest T~dwarfs.

The resolved $(CH_4s-H)$ colors provide more accurate estimates of the
spectral types than the broadband colors, as the $H$-band methane
absorption correlates well with overall near-IR spectral type (\eg,
Figure~2 of \citealp{2005AJ....130.2326T}).
We convert these colors to near-IR spectral type using the polynomial
fit from \citet{liu08-2m1534orbit}:
\begin{eqnarray}
SpT          & = &  19.40 - 19.698 \times (CH_4s-H) - 8.3600 \times (CH_4s-H)^2
\end{eqnarray}
where $SpT=20$ for T0, =~21 for T1, etc. The RMS scatter about the fits
is 0.3~subclasses. The fit was originally derived for objects from T0 to
T8, but as discussed below the more recently discovered T9 objects are
still well-described by this relation.

The observed $(CH_4s-H)$ colors for \twomassbinAB\ give spectral types
of T1.6~$\pm$~0.9 and T9~$\pm$~2 for components A and~B, respectively,
where the spectral type uncertainties come from propagation of the
errors in the colors. Note that the $(CH_4s-H)$ color of component~B
($-0.8\pm0.3$~mag) is bluer than any of the objects used to define the
polynomial fit, so the spectral type estimate represents a small
extrapolation of the relation. For comparison, the colors of the T8.5
dwarfs ULAS~1238+0953 and ULAS~2146$-$0010 are $-$0.65 and $-$0.63~mags,
respectively, and those of the T9~dwarfs ULAS~0034$-$0052,
CFBDS~0059$-$0114, and ULAS~1335+1130 are $-$0.56, $-$0.63, and
$-$0.66~mags, respectively, as synthesized from their published near-IR
spectra \citep{2007MNRAS.381.1400W,
  2008A&A...482..961D,
  2008MNRAS.391..320B,burningham08-T8.5-benchmark}.  (These very
late-T dwarfs are not included in the Liu \etal\ fit, but their
inclusion would make no significant difference in the polynomial
relation.)
Overall, component~B shows $H$-band methane absorption comparable to
the latest known T~dwarfs.


\subsection{Composite Spectra and Resolved Spectral Types \label{sec:spectra}}

We modeled the integrated-light spectrum of \twomassbin\ as the
composite sum of two template T~dwarfs, as an additional means to
determine the resolved spectral types.  The templates were chosen from
available near-IR spectra for dwarfs that have similar colors to those
of 2MASS~J1209$-$1004~A and~B. We produced various fits using
SDSS~J0758+3247 (T2; \citealp{2004AJ....127.3553K}),
SDSS~J1254$-$0122 (T2; \citealp{2000ApJ...536L..35L}) and
SDSS~J1521+0131 (T2; \citealp{2004AJ....127.3553K}) as
templates for the component~A and Gl~570D (T7.5;
\citep{2000ApJ...531L..57B}), 
2MASS~J0415$-$0935 (T8; \citealp{burg01}), ULAS~J1017+0118
(T8p; \citealp{2008MNRAS.391..320B}) and ULAS~J1315+0826
(T7.5; \citealp{2008MNRAS.390..304P}) as templates for component~B.
The $JHK$ colors of the template dwarfs and the resolved binary are
within 0.2~mag for the component~A and 0.4~mag for the component~B
(whose observed photometry is much more uncertain).

Before combining each template pair, we scaled each template spectrum
to a distance of 10~pc. Trigonometric parallaxes are known for
SDSS~J1254$-$0122 and 2MASS~J0415$-$0935. For the other 
dwarfs distances were estimated using the relationships between
absolute magnitudes and spectral type from
\citet{2006astro.ph..5037L}.  We explored fits using both the "bright"
and "faint" relations, which differ significantly for early-type T
dwarfs. For each pair, after shifting the flux to 10pc, the flux at
each wavelength was summed, and the result compared to the observed
unresolved spectrum for \twomassbinAB. There is no parallax
measurement for \twomassbinAB, and so we shifted the observed spectrum
to that of the synthesized spectrum, determined the implied distance,
and compared the resulting spectra.

\begin{figure}
\centerline{\includegraphics[width=4.5in,angle=0]{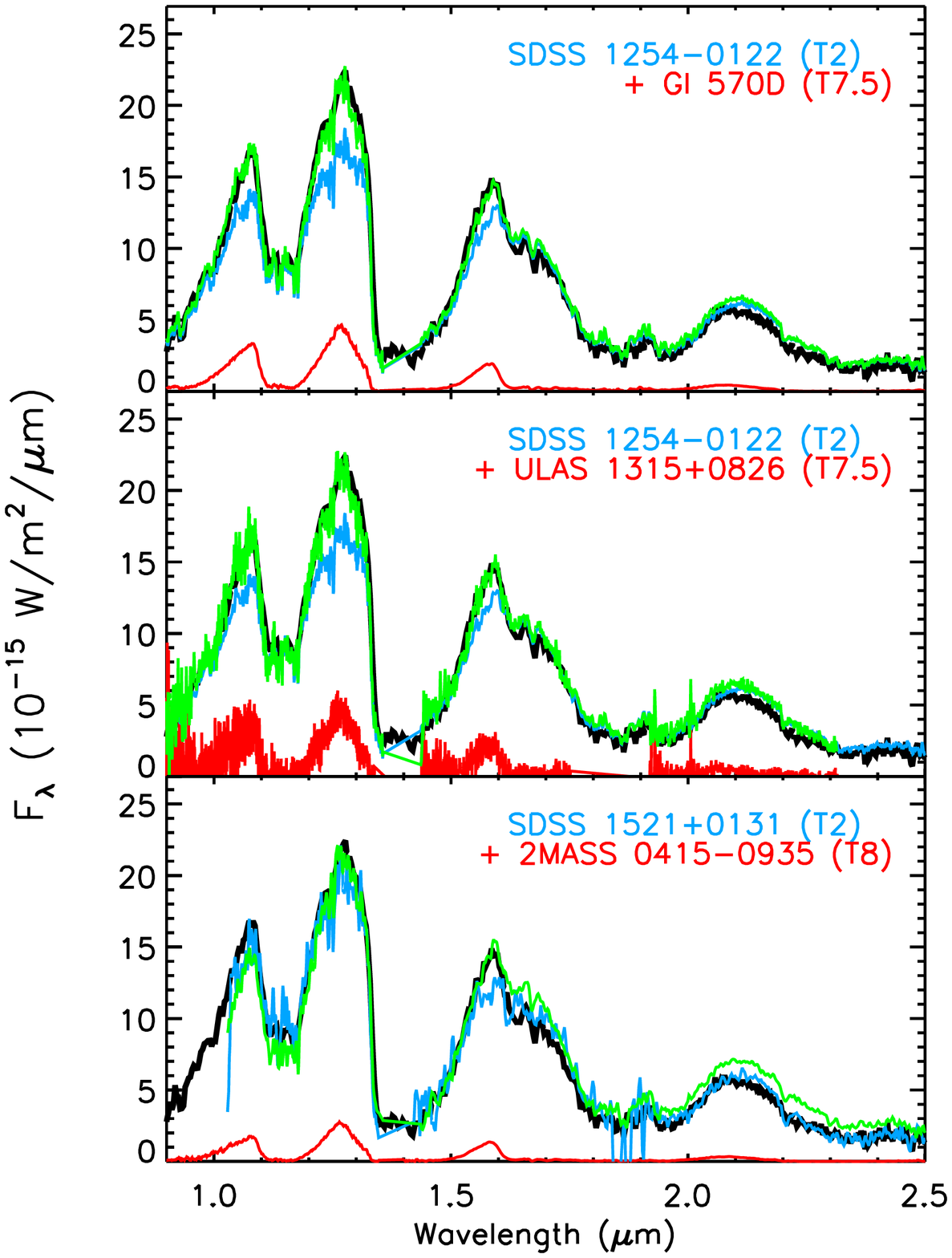}}
\vskip 1ex
\caption{\normalsize The integrated-light near-IR spectrum
  of \twomassbinAB\ (heavy black line; \citealp{2004AJ....127.2856B})
  modeled as the sum of two T~dwarfs (green lines): one early-T
  template (blue lines) and one late-T template (red lines).    The
  templates are chosen based on their similar $JHK$ colors to the two
  components of \twomassbinAB. The agreement between the observed
  spectrum and the modeled blends is 
  good, meaning that the spectral types inferred for \twomassbin{A}
  and~B from the near-IR photometry are consistent with the
  integrated-light spectrum. \label{fig:blends}}
\end{figure}

In the end, model composite spectra composed of a T2~primary and
T7.5~secondary showed good agreement with the observed
integrated-light spectrum (Figure~\ref{fig:blends}). Any combination
using the peculiar T8~dwarf ULAS~J1017+0118 failed to reproduce the
observed strength of the \meth\ feature at 1.6~\micron. The best fits
were obtained using the T2~dwarf SDSS~J1254$-$0122 in combination with
either of the T7.5~dwarfs Gl~570D or ULAS~J1315+0826, or using the
T2~dwarf SDSS~J1521+0131 in combination with the T8~dwarf
2MASS~J0415$-$0935. For the last combination to work, the distance for
SDSS~J1521+0131 must be derived using the "faint" Liu \etal\
photometric relation. The implied distance for \twomassbinAB\ is 23~pc
from the SDSS~J1254$-$0122 fits, or 18~pc from the SDSS~J1521+0131
fit.

Combining the results from the $JHK$ and $(CH_4s-H)$ colors and this
spectral modeling, we adopt a spectral type of T2~$\pm$~0.5 for
\twomassbin{A}, as all the available estimates are in good agreement.

The typing for component~B is more challenging, given its larger
photometric errors and its relative faintness (and thus modest
contribution to the integrated-light spectrum). In the spectral
modeling, T7.5 templates provided the best matches. Our attempts to use
T8~dwarfs failed to produce a good match, though this was possibly
limited by the small number of templates with comparably blue near-IR
colors as \twomassbin{B} (2MASS~J0415$-$0935 and ULAS~J1017+0118). The
polynomial fits for absolute magnitude as a function of spectral type
from \citet{2006astro.ph..5037L} give $M(J$) = \{15.1,
  15.5, 15.9, 16.3\}~mag, $M(H)$ = \{15.5, 15.9, 16.3, 16.7\}~mag, and
  $M(K)$ = \{15.6, 16.0, 16.4, 16.7\}~mag for near-IR spectral types of
  T6.5, T7, T7.5, and T8, respectively. Based on the photometric
  distance to component~A (\S~\ref{sec:distance}), the absolute
  magnitudes of component~B are $M(J, H, K) = 15.7\pm0.4,
  16.5\pm0.5,
  16.9\pm0.7$~mag, in good agreement with T7.5. Given the concordance
  between this and the spectral modeling, we adopt a spectral type
  estimate of T$7.5\pm0.5$ for component~B.


\subsection{T3 Near-IR Spectral Type Standard}

\citet{2005astro.ph.10090B} chose \twomassbin\ as the
  primary T3 spectral standard for near-IR classification of T~dwarfs.
  Its binary nature now lessens its utility for this purpose, especially
  at the bluer wavelengths where the contribution of the secondary to
  the integrated-light spectrum is greater.
  The secondary standard proposed by Burgasser \etal\ was
  SDSS~J1021$-$0304AB, which also has been resolved into a binary
  (T1+T5; \citealp{2006ApJS..166..585B}). For both sources, their binary
  nature make them unappealing as the T3 spectral standard, since the
  secondary contributes significantly to the integrated near-infrared
  light (\eg, Figure 3).\footnote{The mid-infrared regime becomes
    increasingly dominant for cooler T dwarfs (e.g. Figure~3 of Cushing
    \etal\ 2006) and so the T7.5~\twomassbin{B}\ is also expected to
    make a significant contribution to the integrated \twomassbin\ flux
    at $\sim$10~\micron.} Furthermore the primary of neither system is
  of T3 type.
Options for a different spectral standard are limited, due to the modest
number of T3 dwarfs known. At the relatively brighter magnitudes, the
choices are SDSS~J120602.51+281328.7 (T3; $J=16.5$~mag) and
SDSS~J141530.05+572428.7 (T3$\pm$1; $J=16.7$), both of which were
identified and classified by \citet{chiu05}. SDSS~J1206+2813 appears to
be single in 0.06\arcsec\ FWHM images obtained by us with Keck LGS (Liu
\etal, in prep.) while no high angular resolution imaging is available
yet for SDSS~J1415+5724. As the individual
  spectral indices for SDSS~J1206+2813 are more consistent with a T3
  typing and it appears to be single, we suggest that it be used as the T3
  spectral standard.


\subsection{Photometric Distance \label{sec:distance}}

With an estimated spectral type of T$2.0\pm0.5$ for component~A, we
use the \citet{2006astro.ph..5037L} polynomial fits of absolute
magnitude as a function of spectral type to determine a photometric
distance.  We use the $J$-band photometry; the $H$ and $K$-band data
produce nearly identical results but with somewhat larger
uncertainties.  Errors in the photometric distance include the
uncertainties in the apparent magnitude (0.05~mag), the uncertainty in
$J$-band absolute magnitude arising from the spectral type uncertainty
(0.05~mag), and the intrinsic dispersion about the fit (0.4~mag).  Liu
\etal\ provide both a ``bright'' and a ``faint'' relation, depending
on the inclusion or exclusion of candidate overluminous objects of
early/mid-T~spectral type, respectively.  The two relations give
consistent answers within the errors, $24\pm4$ and $18\pm4$~pc,
respectively.  We average to arrive at a final photometric distance of
$21\pm4$~pc.\footnote{In a similar fashion, we find a photometric
  distance to component~B of $19\pm5$~pc based on its type of
  T$7.5\pm0.5$.  This value is consistent with the distance derived
  component~A, but it cannot be considered truly an independent
  measurement, since we use the photometric distance in estimating the
  spectral type for component~B in \S~\ref{sec:spectra}.}

Since the spectral templates used for modeling the composite spectra
have distance determinations (\S~\ref{sec:spectra}), the scaling
factor used to match to \twomassbin{AB} also provides a distance
estimate to the binary, with a value of 18 or 23~pc
(\S~\ref{sec:spectra}).  An error estimate from this approach is
uncertain.  The T2~dwarf SDSS~J1254$-$0122 \citep{2000ApJ...536L..35L}
provides a well-matching template for component~A and has parallax
with an uncertainty of 4\% \citep{2004AJ....127.2948V}, which
represents the very smallest possible uncertainty.  At the other
extreme, the photometric relations of \citet{2006astro.ph..5037L} have
a RMS scatter about the fit of 15--20\% in distance.  Adopting a
distance error of 10\% is a reasonable compromise; for instance, Liu
\etal\ demonstrate that fitting the composite spectrum of the T1+T6
binary $\epsilon$~Ind~Bab produces a distance estimate that agrees to
10\% with the measured parallax.  Thus, the distance estimates from
the spectral modeling agree with the photometric distance of
$21\pm4$~pc.  We conservatively adopt the latter in our subsequent
analysis.\footnote{The slightly larger distance from the spectral
  modeling results from the fact that SDSS~J1254$-$0122 is slightly
  brighter than the values from the adopted Liu \etal\ polynomial
  fits.  In fact, it has been suggested this system may be a binary
  based on this modest overluminosity \citep{2006astro.ph..5037L,
    burg2006-lt}, though it has not been resolved as such in high
  angular resolution imaging from \HST\ or Keck LGS AO
  (\citealp{2006ApJS..166..585B}; Liu \etal\, in prep.)  and modeling
  of its integrated-light spectra suggests the system is more likely
  be a relatively young, single object as opposed to a near-equal mass
  binary \citep{stephens09-irac}.}


\subsection{Bolometric Luminosities \label{sec:lbol}}

\subsubsection{Combined System}

There is no published value for the integrated-light bolometric
luminosity of \twomassbin, so we computed it directly from its near-IR
spectrum \citep{2004AJ....127.2856B}, flux calibrated with the
\citet{chiu05} photometry, and combined with mid-IR photometry.
For the latter, IRAC photometry is available in the
  \Spitzer\ archive, obtained as part of the Cycle~1 GTO program~35 on
  UT 2005~June~13. Exposure times were 30~s, and a 5-position small
  dither pattern was used. The binary is unresolved in the IRAC images,
  and an 8" aperture was used for the photometry. We determined
  magnitudes of $[3.6]=14.02\pm0.004$, $[4.5]=13.49\pm0.004$,
  $[5.8]=13.33\pm0.016$, and $[8.0]=13.06\pm0.06$~mags. A 0.03~mag error
  should be added in quadrature to the quoted random errors to account
  for systematic effects.

To derive the integrated-light \Lbol, we numerically integrated the
near-IR spectrum and the mid-IR photometry, interpolating between the
gaps in the data, extrapolating at shorter wavelengths to zero flux at
zero wavelength, and extrapolating beyond 4.5~\micron\ assuming a
blackbody. We determined the luminosity error in a Monte Carlo fashion,
by adding randomly drawn noise to our data over many trials and
computing the rms of the resulting luminosities. In this process, we
accounted for the noise in the spectrum, the errors in the MKO
photometry used to flux-calibrate it, and the errors in our mid-IR
photometry estimates. Before accounting for the error in the photometric
distance, we found a total bolometric luminosity of $\log(\Lbol/\Lsun) =
-4.61\pm0.02$~dex. With the 20\% uncertainty in the photometric distance
included, the \Lbol\ uncertainty becomes 0.15~dex.

\subsubsection{Individual Components}

To compute the bolometric luminosities (\Lbol) for each component, we
apply bolometric corrections to the resolved near-IR photometry based on
the compilation in Appendix~\ref{sec:bolcorr}. Despite the relatively
larger RMS about the $BC$ fit at $J$-band, the photometry of the
component~B is most accurately determined in this bandpass, so we use it
to to determine \Lbol\ and associated quantities. The results are given
in Table~\ref{table:resolved}. The \Lbol\ uncertainties are composed of
the uncertainties from the apparent magnitudes, from the bolometric
corrections arising from the spectral type uncertainty, the intrinsic
dispersion about our polynomial fits, and the photometric distance. The
last item dominates the uncertainties in the \Lbol\ values.
(However, the luminosity {\em ratio} of two components is known to much
higher precision than the individual \Lbol\ values, since the ratio is
independent of the distance uncertainty.) As a consistency check, we
note that the sum of the individual component luminosities derived from
$BC_J$ ($-4.51\pm0.18$~dex) agrees well with the direct computation from
the integrated light spectrum ($-4.61\pm0.15$~dex).


\subsection{Physical Properties from Evolutionary Models \label{sec:properties}}

\begin{figure}
\vskip -0.5in
\hskip 0.3in
\includegraphics[width=3.6in,angle=90]{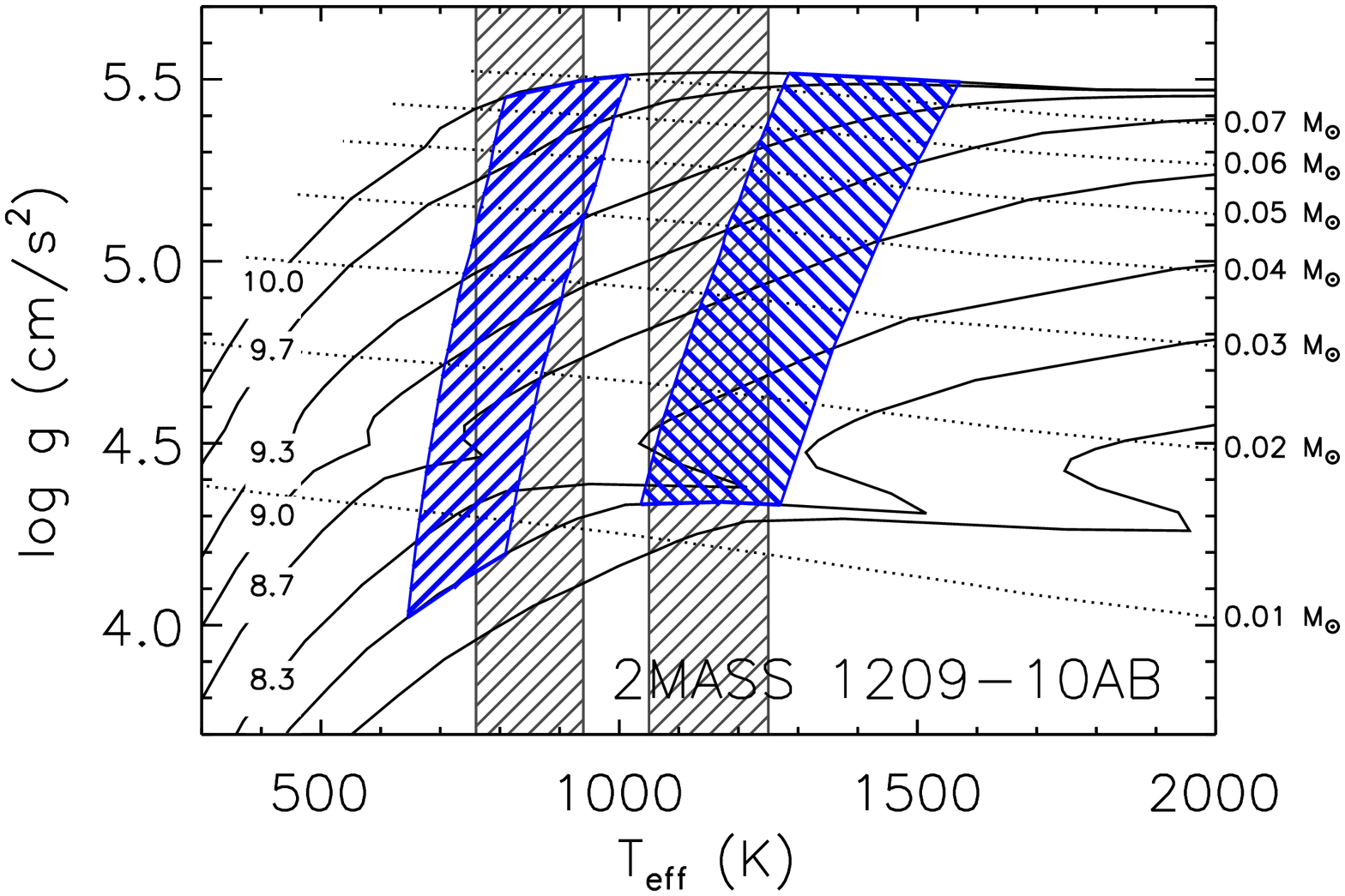}
\vskip -0.2in
\hskip 0.3in
\includegraphics[width=3.6in,angle=90]{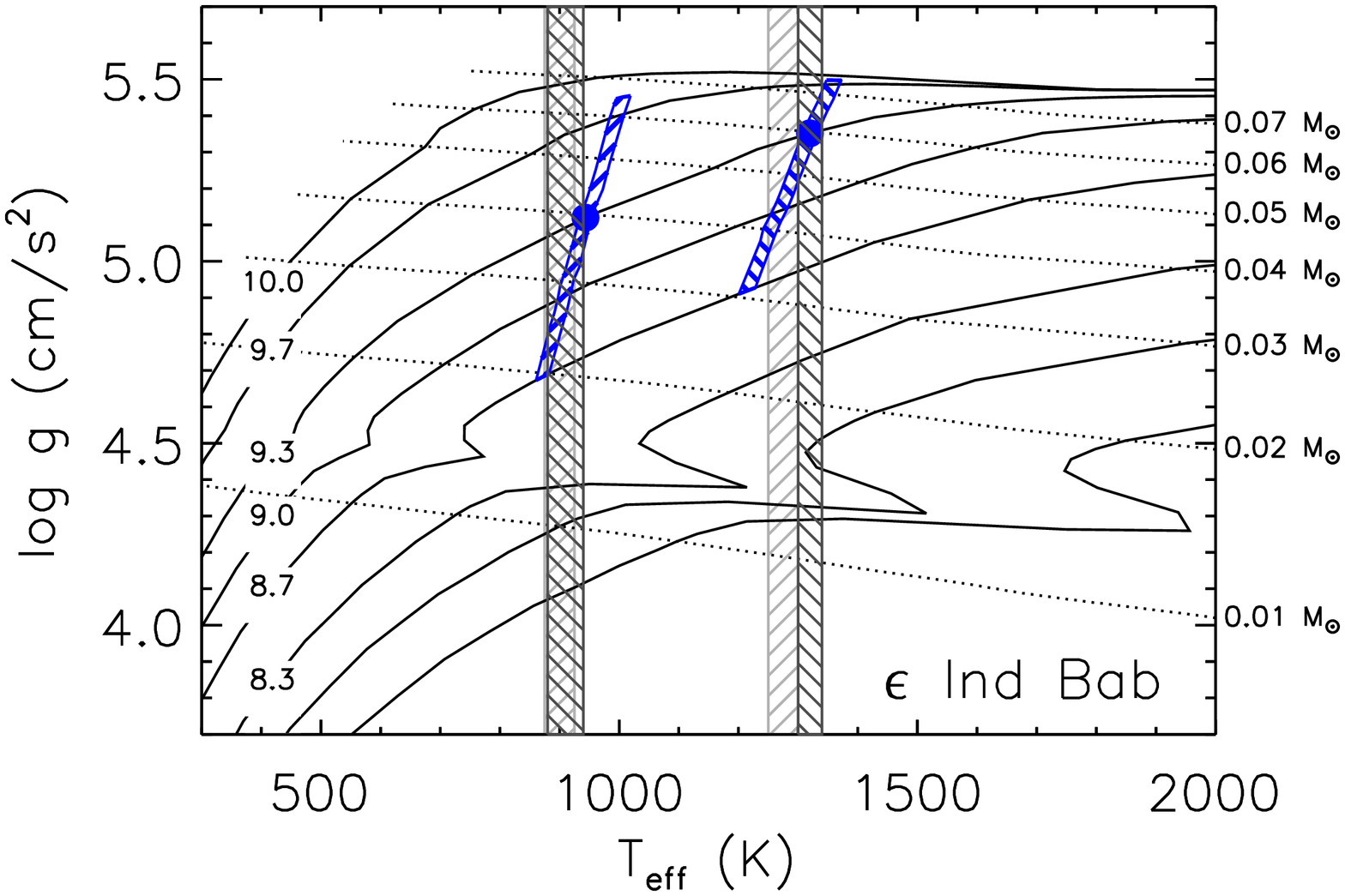}
\vskip -3ex
\caption{\normalsize Inferred \Teff\ and \logg\ for the components of
  \twomassbinAB\ and \eIndBab\ based on solar-metallicity evolutionary
  models by \citet{1997ApJ...491..856B}. The blue-colored hatched
  regions show the constraints from the observed \Lbol's of the binary
  components and the inferred system ages, with the A~component for each
  system being the region on the right (\ie, the hotter object).  For
  \twomassbinAB, we plot a generic age range of 0.1--10~Gyr. For 
  \eIndBab, we plot an age range of 0.5--7.0~Gyr, with the filled circle
  representing 2~Gyr. The uncertain \Lbol's of \twomassbinAB\ are
  reflected in its relatively wide hatched regions; these can be improved
  with a parallax determination for the system.
  The solid lines are isochrones from 0.1 to 10~Gyr, labeled by the
  logarithm of their age; the two unlabeled isochrones are 0.05 and
  0.1~Gyr. Dotted lines represent iso-mass models, labeled on the right
  of the plot in units of \Msun. 
  The vertical grey regions show the
  \Teff's for the two components derived from model atmosphere fitting;
  for \eIndBab, we show the \Teff\ values from both
  \citet{2009ApJ...695..788K} (cooler values, in light grey) and
  \citet{2010A&A...510A..99K} (hotter values, in dark grey).
  \label{fig:teff-logg}}
\end{figure}

With the \Lbol\ determinations for the individual components and an
assumed age, evolutionary models can then fully determine the physical
properties of the two components \citep[e.g.][]{2000ApJ...541..374S}.
Figure~\ref{fig:teff-logg} shows the results graphically using the
\citet{1997ApJ...491..856B} evolutionary models, where we have adopted a
generic age range of 0.1--10~Gyr for the \twomassbinAB\ system. The
figure indicates the complete possible set of \{\Teff, \logg\} values
for the two components, based on our \Lbol\ determinations. 

Table~\ref{table:evolmodels} summarizes the physical properties derived
from this approach, for a range of assumed ages.
Note that for computing relative quantities between the components (\eg,
luminosity ratio and mass ratio), we take care to account for the
covariance of the measurements due to the fact that the uncertainty in
the distance is common to both components, and thus uncertainties in
these relative quantities are smaller than for the absolute value of
these quantities for the individual components. To do this, we draw
values for all the relevant measurements in a Monte Carlo fashion,
keeping track of all the calculations of the physical parameters in
every trial, and then compute the final result and RMS from the ensemble
of trials.

To compute the estimated orbital period of the system, following
\citet{1999PASP..111..169T}, we assume random viewing angles and a
uniform eccentricity distribution from $e = 0-1$. This gives a
multiplicative correction factor of 1.10$^{+0.91}_{-0.36}$ (68.3\%
confidence limits) for converting the projected separation into a
semi-major axis. The estimated orbital periods range from 16--26~years
for ages of 0.5, 1.0, and 5.0~Gyr, albeit with significant
uncertainties. Dynamical masses for ultracool binaries are possible
from orbital monitoring covering $\gtrsim$30\% of the orbital period
\citep[e.g.][]{2004A&A...423..341B, liu08-2m1534orbit}.  Thus,
\twomassbinAB\ warrants continued high angular resolution imaging to
determine its visual orbit, combined with a parallax measurement to
compute its total mass.


\section{Discussion: The Isochrone Test \label{sec:isochrone}}

\begin{figure}
\vskip -1in
\centerline{\includegraphics[width=4.5in,angle=90]{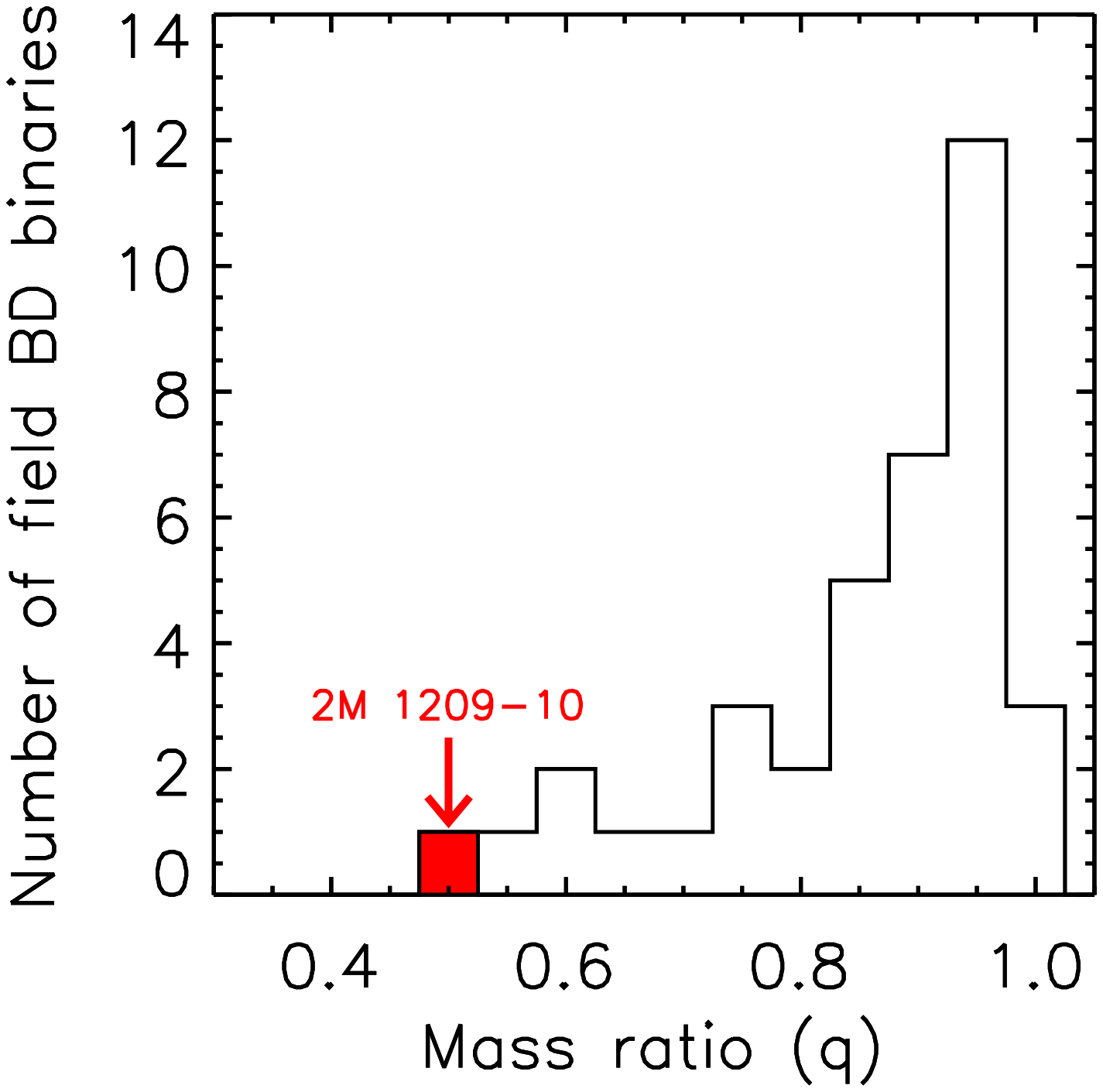}}
\caption{\normalsize Mass ratios for all known field brown dwarf
  binaries, based on our compilation given in Table~\ref{table:binaries}
  and described in Appendix~\ref{sec:binaries}. The mass ratio for
  \twomassbinAB\ from our calculations is shown as the filled color bar,
  demonstrating the rarity of such an unequal mass ratio among the field
  population. (SDSS~J141624.08$+$134826.7AB is not plotted, given its
  very unusual properties relative to all other known systems and the
  large systematic uncertainty in its mass ratio; see
  Appendix~\ref{sec:binaries}.)
 \label{fig:massratio}}
\end{figure}

The \twomassbinAB\ system is a rare example of an ultracool binary with
a highly unequal mass ratio ($q\approx0.5$), as most ultracool binaries
are composed of nearly equal components (Table~\ref{table:binaries}).
This novelty is highlighted in Figure~\ref{fig:massratio}, which shows
the estimated mass ratios of all known field brown dwarf binaries, where
we choose field systems in which the primary has an estimated spectral
type of L4 or later as a proxy for the stellar/substellar boundary
\citep{1999ApJ...519..802K}. As explained below, we specifically focus
on ultracool binaries composed of two brown dwarfs, as opposed to those
that contain one or two very low-mass stars.

Under the conservative assumption of coevality, we can use
\twomassbinAB\ to test current theoretical models of brown dwarfs, by
assessing whether the two components lie along a single isochrone on the
Hertzsprung-Russell (H-R) diagram. We essentially treat the system as a
star cluster composed of two members, with an unknown age but requiring
that it is the same for components~A and~B. Hereinafter, we refer to
this approach as the ``isochrone test.''

Placement on the H-R diagram requires \Lbol\ and \Teff. The former can
be well-determined for ultracool dwarfs (Appendix~\ref{sec:bolcorr} and
references therein). However, robust temperature determinations are
still an active area of investigation
\citep[e.g.][]{2007arXiv0711.0801C}. Atmospheric models for L~and
T~dwarfs have not been confirmed by direct measurement of brown dwarf
radii, only checked for consistency against evolutionary models using
single brown dwarfs that are companions to stars of known age
\citep[e.g.][]{2006ApJ...647..552S, 2008arXiv0804.1386L} and the few
brown dwarf binaries with dynamical mass determinations
\citep[e.g.][]{liu08-2m1534orbit, 2008arXiv0807.2450D,
  2009ApJ...699..168D}.\footnote{Of course, for this purpose we cannot
  use temperatures that are derived from the evolutionary models
  themselves \citep[e.g.][]{gol04}, where the observed \Lbol, an assumed
  age, and the evolutionary-model-derived radii are used to compute
  \Teff. If we did, then any information gleaned from the H-R diagram
  positions would automatically agree with the evolutionary models, \ie,
  this would partly be circular reasoning.} Modelling low-temperature
atmospheres is challenging. The molecular opacity line list for CH$_4$,
a species important for both the L and T dwarfs, is incomplete or
non-existent below 1.6~\micron. The FeH line list is incomplete in the
1.2--1.3~\micron\ region, which is important for L dwarfs. NH$_3$ is
important for very late-type T dwarfs, but there are no opacities
calculated below 1.4~\micron. Also, calculation of the very strongly
pressure-broadened \ion{K}{1} resonance doublet (0.78~\micron) is
difficult, and hence models do not currently reproduce the far-red
region of T dwarf spectra very well. Finally, the treatment of the
grains condensates is difficult, which impacts the 1.0--1.6~\micron\
region of $\approx$L3~to $\approx$T2~types.

Thus, our current ability to carry out the isochrone test represents a
test of the {\em joint} accuracy of evolutionary and atmospheric models,
not of either specific class of models.\footnote{One
    limitation on the isochrone test is that the currently available
    evolutionary models do not typically incorporate the same set of
    model atmospheres used in fitting observed spectra.  However,
    \citet{2000ApJ...542..464C} show that the predicted luminosity
    evolution is mostly robust against the adopted model atmosphere.}
The fact that the isochrone test relies on components of very different
masses also may make it a useful signpost for mass-dependent problems,
which would not be revealed when studying the (much more common)
nearly-equal mass binaries. Of course, the coevality requirement is
widely applied in studies of star clusters. Its particular
characteristic in the case of substellar binaries arises from the fact
that brown dwarfs steadily cool over their lifetime, and thus the
isochrones for substellar objects are more dispersed on the H-R diagram
compared to stellar isochrones.\footnote{The isochrone test described
  here is conceptually identical to some previous studies of young
  star-forming regions. Although the underlying physical processes are
  different for pre-main sequence stars and field brown dwarfs, the
  luminosities of both classes of objects are strongly dependent on age.
  For young regions, coevality has been used to check the accuracy of
  evolutionary models and the spectral type-to-effective temperature
  scale, using multiple systems \citep[e.g.][]{1994ApJ...427..961H,
    1999ApJ...520..811W, 2008AJ....135.1659S, 2009ApJ...704..531K} and
  entire clusters \citep[e.g.][]{1999ApJ...525..466L}. Note for such
  young objects, a complicating factor is the possible impact of
  accretion on the H-R diagram positions \citep{2009ApJ...702L..27B}, an
  effect which is not relevant for old (field) brown dwarfs.}

\subsection{\twomassbinAB \label{sec:isochrone-2m1209}}

We now subject \twomassbinAB\ to the isochrone test. A parallax
measurement is not available, so we use the photometric distance
estimate (\S~\ref{sec:distance}) --- this is not a limiting factor in
the analysis. For the effective temperatures of each component, we
consider previous analysis of the spectra for similarly typed T dwarfs.

Three T2~dwarfs have been analyzed in such a fashion: SDSS~J0758+3247
(1100--1200~K; \citealp{stephens09-irac}), SDSS~J1254$-$0122
(1100--1200~K; \citealp{stephens09-irac, 2007arXiv0711.0801C}), and
HN~Peg~B (1115~K; \citealp{2008arXiv0804.1386L}).\footnote{The first two
  objects were analyzed with direct fitting of model atmospheres to
  their observed spectra. For HN~Peg~B, the model atmosphere matching
  the \Teff\ derived from evolutionary models was shown to be consistent
  with the observed flux-calibrated spectrum.} The model atmospheres
were spaced by 100~K (except for HN~Peg~B), and the estimated systematic
uncertainties are $\approx$100~K due to, \eg, flux calibration, choice
of wavelength region for fitting, and weighting scheme. Thus we adopt
$1150\pm100$~K as the temperature for \twomassbin{A}. No~T1 or T3~dwarfs
have been subjected to such analysis, but the \Teff\ fitted for
early/mid-T~dwarfs are all relatively constant
\citep[e.g.][]{stephens09-irac}, in accord with the \Teff\
  measurements derived from \Lbol\ data that the L/T transition for
  field objects occurs at nearly constant temperature
  \citep[e.g.][]{gol04}.

  Four T7.5~dwarfs have been analyzed to date: HD~3651B (820--830~K;
  \citealp{2007arXiv0705.2602L}),
  2MASS~J1114$-$2618 (725--775~K;
  \citealp{2007arXiv0705.2602L}), Gl~570~D
  (800-821~K; \citealp{2006ApJ...647..552S}), and 2MASS~J1217$-$0311
  (850--950~K; \citealp{2006astro.ph.11062S}).
  To encompass the potential uncertainty in the spectral type of
  component~B, we also consider the T8~dwarf 2MASS~J0415$-$0935
  (725--775~K, $710\pm40$~K; \citealp{2006astro.ph.11062S,
    2008ApJ...689L..53B}).\footnote{For Gl~570D and 2MASS~J0415$-$0935,
    the model atmosphere matching the \Teff\ derived from evolutionary
    models was shown to be consistent with the observed flux-calibrated
    spectrum, as opposed to directly fitting model atmospheres to the
    spectra.}
  No T7~dwarfs have independently determined
  \Teff's.\footnote{\citet{2006ApJ...639.1095B} have determined \Teff's
    for a sample of mid/late-T~dwarfs based on multi-band flux ratios
    measured from low-resolution near-IR spectra. However, these are
    calibrated against the \Teff\ and \logg\ of Gl~570D derived directly
    from evolutionary models. And thus these \Teff's cannot be used to
    place objects on the H-R diagram, since this would be circular
    reasoning.}
  Taking the unweighted mean and RMS of the results gives
  $\Teff=800\pm70$~K as our adopted value for \twomassbin{B}.

\begin{figure}
\hskip 0.1in
\includegraphics[width=3in,angle=0]{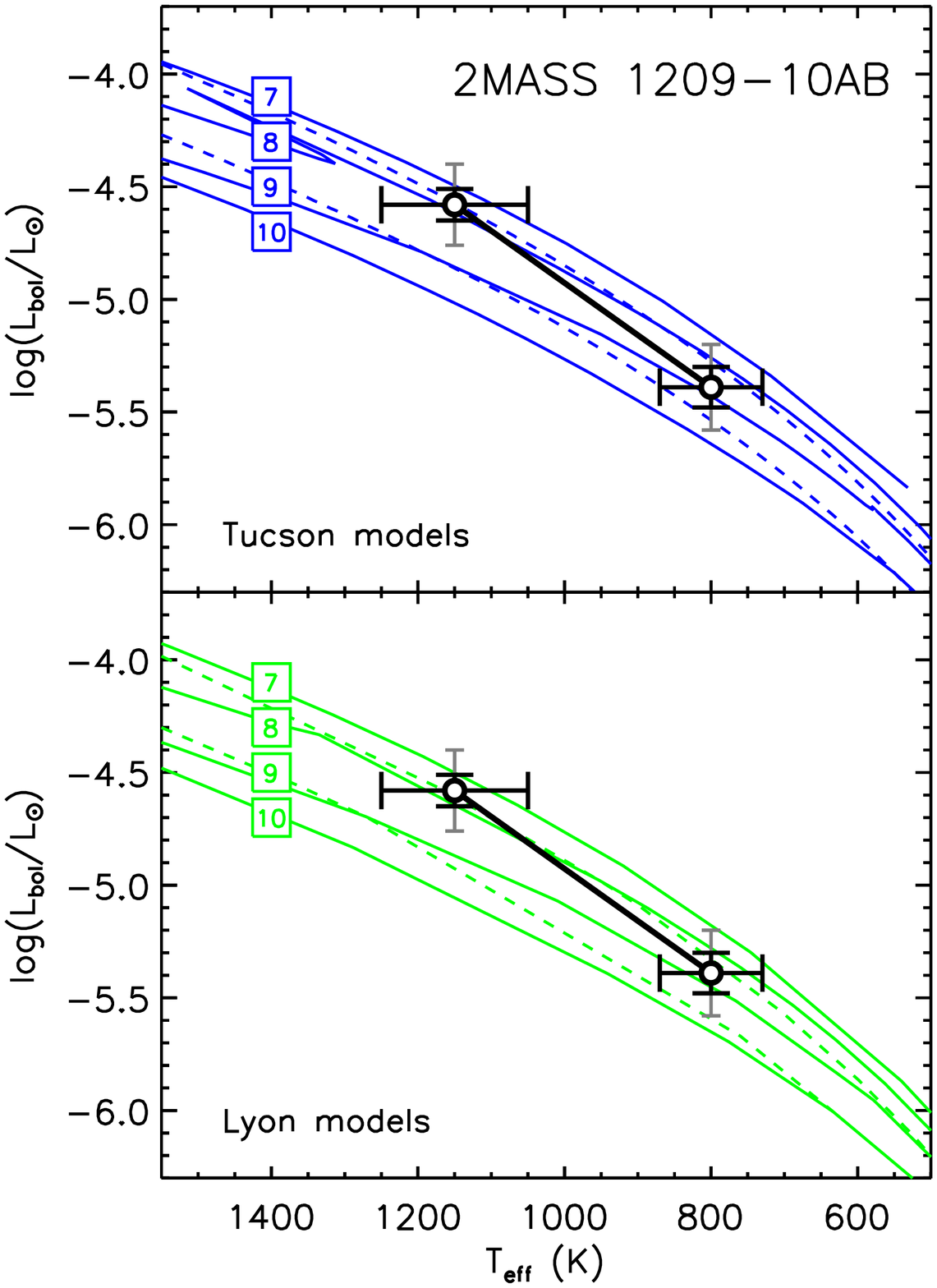}
\hskip 0.5in
\includegraphics[width=3in,angle=0]{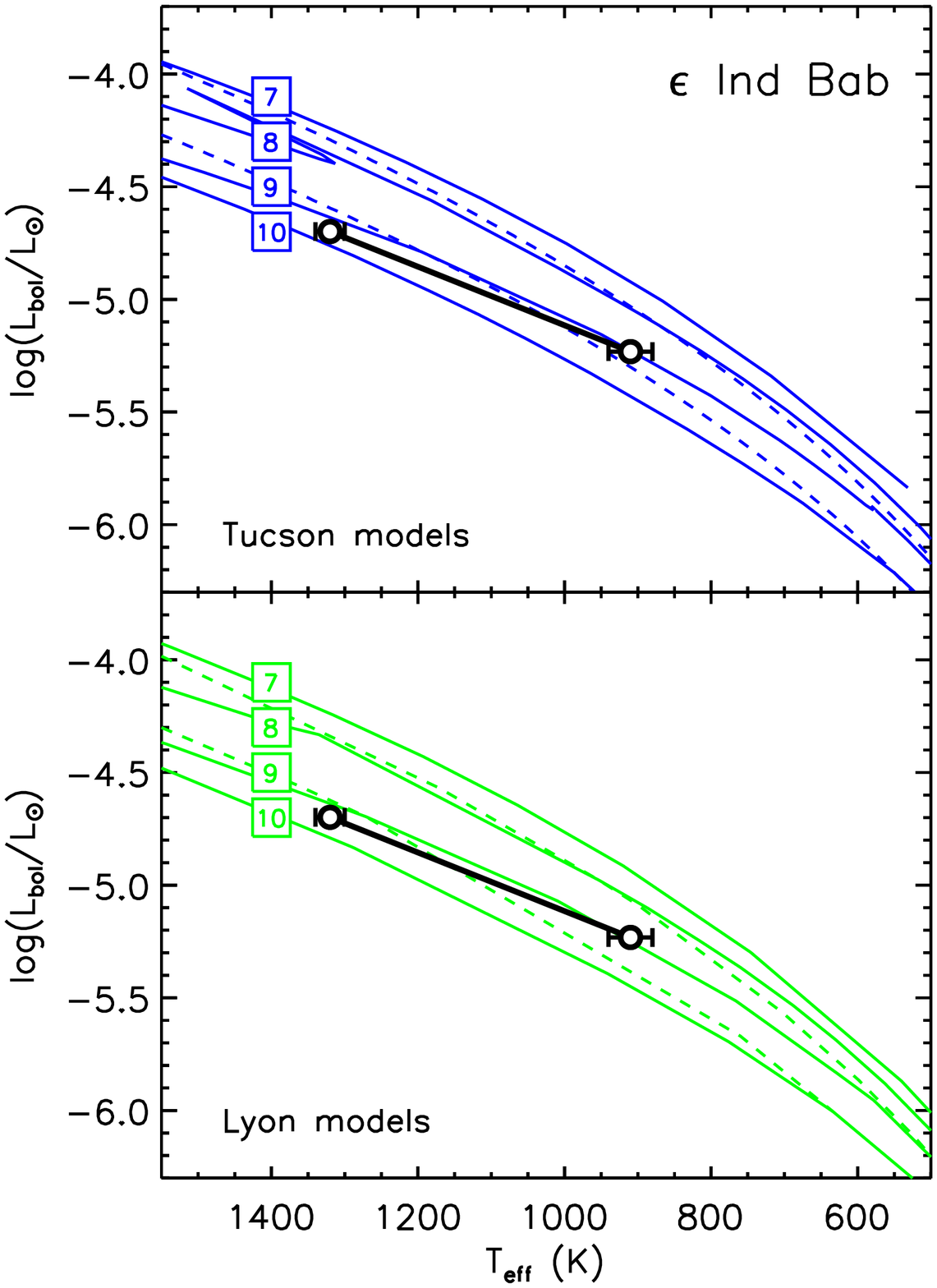}
\vskip 3ex
\caption{\normalsize H-R diagram showing the two components of
  \twomassbinAB\ and \eIndBab\ compared to isochrones and iso-mass
  tracks from the Tucson and Lyon (COND) evolutionary models. The data
  points show the individual components. The thick black diagonal
  connects the two components, showing the isochrone defined by the
  binary system. The isochrones are shown as solid colored lines, with
  the numbers in squares on the left side of the plots giving the
  logarithms of the isochrone ages ($10^7, 10^8, 10^9, 10^{10}$~yr). The
  iso-mass tracks for 0.01~\Msun\ (upper line) and 0.04~\Msun\ (lower
  line) are shown as dashed black lines.  
 For \twomassbinAB, two sets of errors are shown for the $y$-axis: the
  longer, thinner grey error bars show the total error in the \Lbol\
  measurements, and the shorter, thicker black errors bars show the
  errors with the uncertainty in the distance removed, since this
  uncertainty is common to both components.
  For \eIndBab, the plot uses the \Teff\ values from \citet{2010A&A...510A..99K}.
\label{fig:hrd-both}}
\end{figure}

\begin{figure}
\hskip -0.65in
\includegraphics[width=3.5in,angle=90]{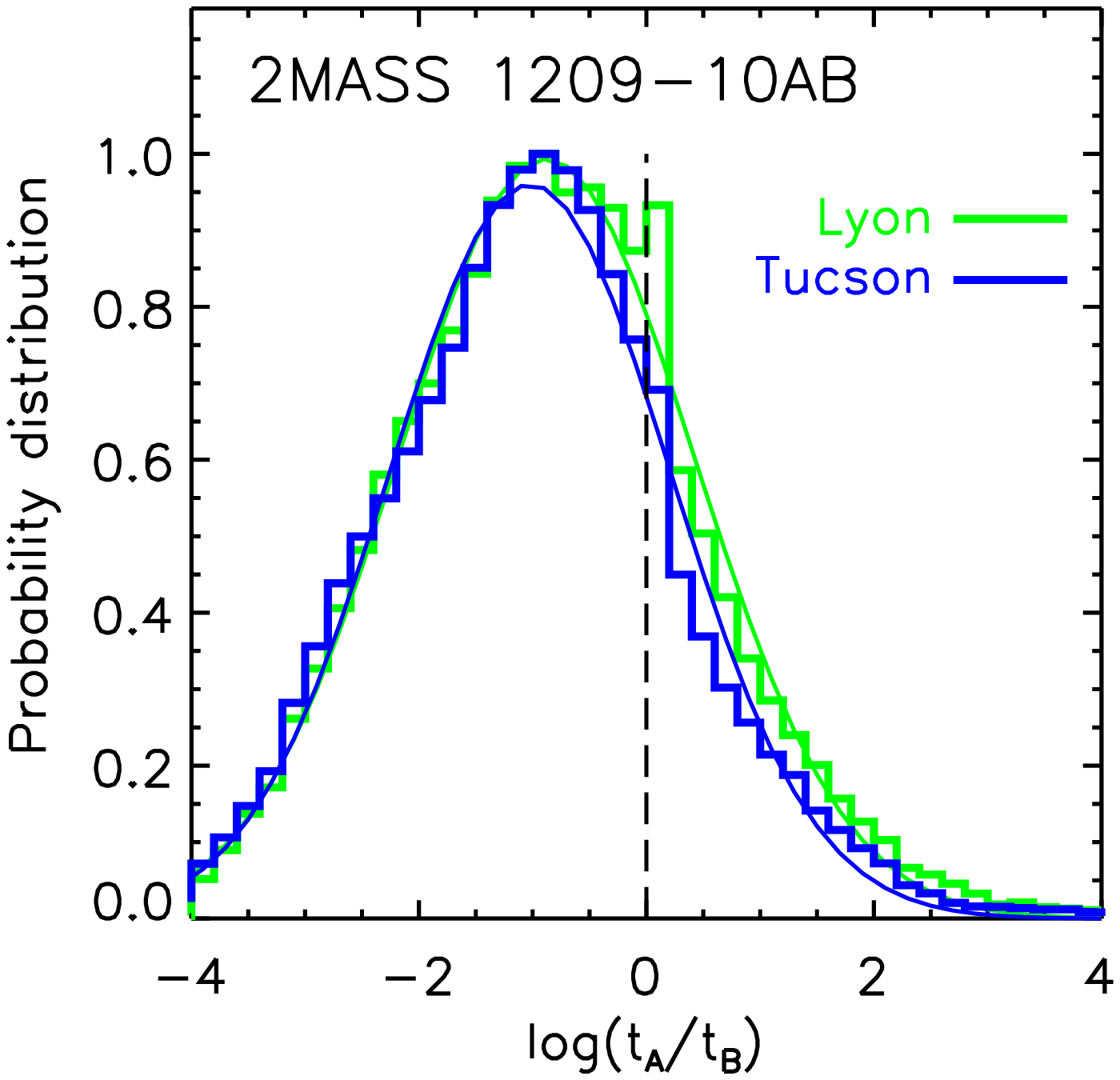}
\hskip -1.5in
\includegraphics[width=3.5in,angle=90]{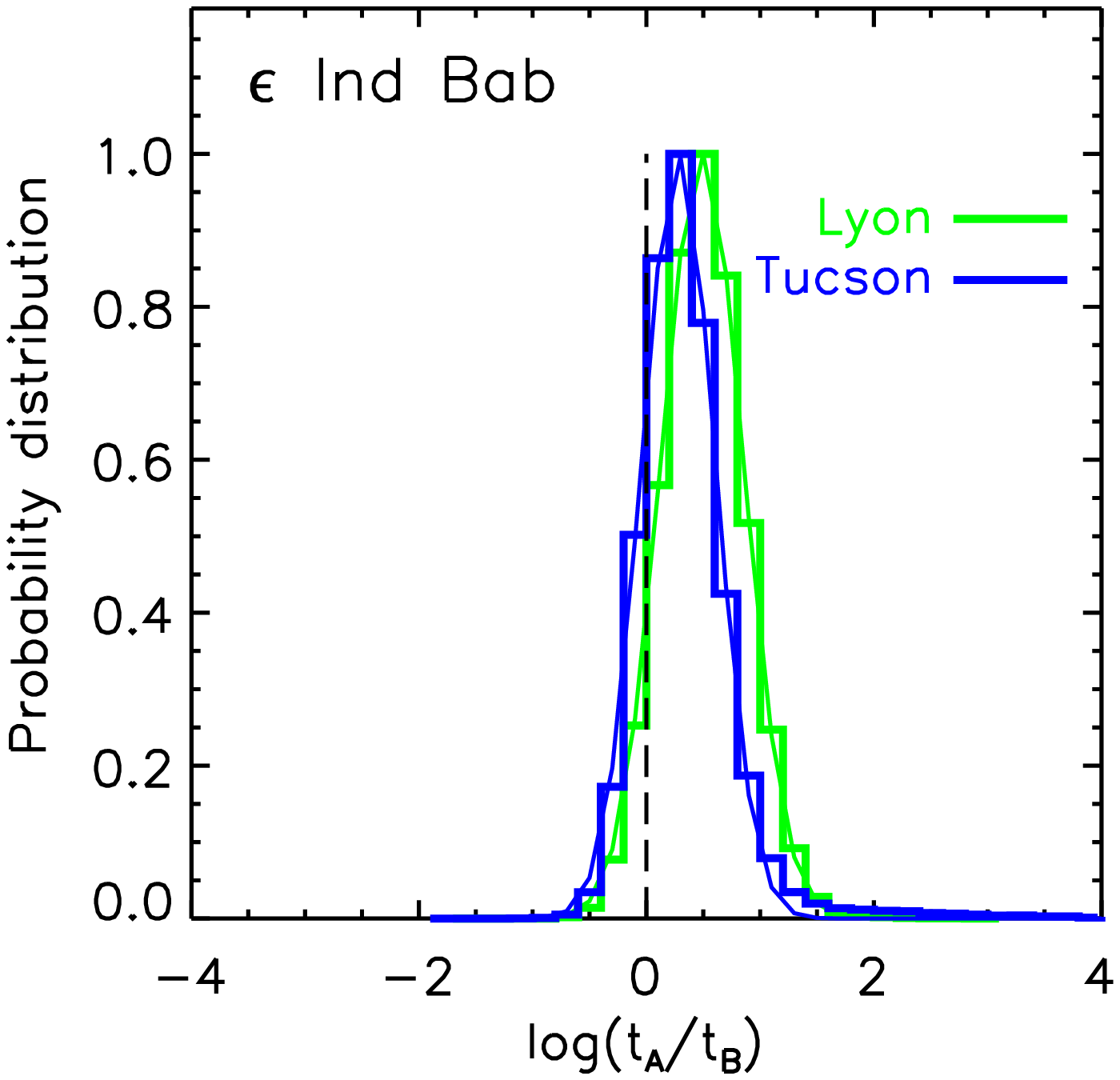}
\caption{\normalsize Probability distribution of the age difference
  between components~A and~B in the \twomassbin{AB} and \eIndBab\
  systems, based on ages inferred from their locations on the H-R
  Diagram (Figure~\ref{fig:hrd-both}) and the Lyon (COND) and Tucson
  evolutionary models. The histogram shows the results from our Monte
  Carlo calculation, and the thin solid line shows the best fitting
  Gaussian to the binned probability distribution. The dashed vertical
  line denotes coevality of the two
  components. \label{fig:hrd-agediff-both}}
\end{figure}

  Figure~\ref{fig:hrd-both} shows \twomassbinAB\ on the H-R diagram. We
  compute the individual component ages based on their H-R diagram
  positions relative to the Lyon/COND \citep{2003A&A...402..701B} and
  Tucson \citep{1997ApJ...491..856B} evolutionary models, using a Monte
  Carlo method to account for the uncertainties. For each component, we
  draw trial values for \Lbol\ and \Teff\ from a normal distribution,
  which are then used with finely interpolated tabulations of the models
  to derive a set of ages. We then take the difference of the component
  ages and compile the distribution resulting from 10$^6$
  trials.\footnote{The two components of \twomassbinAB\ are subject to
    the exact same uncertainty in the photometric distance, and we take
    care to track this aspect properly when drawing the Monte Carlo
    values. The net result is that the uncertainty of the age difference
    between~A and~B is smaller than would be expected simply from the
    uncertainties in the absolute ages of either component.}
  Figure~\ref{fig:hrd-agediff-both} plots the result, which is
  well-described by a gaussian. Table~\ref{table:ages} gives the derived
  ages and age difference. At the 1$\sigma$ level, the models
  successfully indicate that the two components are coeval, with a
  $\log(age)$ difference of $-0.8_{-1.3}^{+1.3}$ and
  $-1.0_{-1.3}^{+1.2}$~dex inferred from the Lyon and Tucson models,
  respectively.

\subsection{\eInd~Bab \label{sec:isochrone-epsInd}}

We also apply the isochrone test to the \eInd~Bab system, which is a
wide (1459~AU) separation binary companion to the K5~V star \eInd{A}
\citep{2003A&A...398L..29S}. The binary is composed of a T1 and a T6
component \citep{2004A&A...413.1029M, 2005astro.ph.10090B}, with a
$K$-band flux ratio of $2.18\pm0.03$~mag. The system is quite similar to
\twomassbinAB, but with more precisely determined physical parameters
and therefore offers a more stringent application of the isochrone
test.\footnote{Table~\ref{table:binaries} indicates that
  2MASS~J22551861$-$5713056AB ($q=0.63$), 2MASS~J05185995$-$2828372AB
  ($q=0.54$), 2MASS~J12255432$-$2739476AB ($q=0.63$) have comparably
  small mass ratios as \twomassbinAB\ and \eInd~Bab. For these other
  binaries, the existing resolved measurements are insufficient to apply
  the isochrone test.}

For the \Lbol\ of the two components, we use the
  results from \citet{2010A&A...510A..99K} of
  $\log(\Lbol/\Lsun)=-4.699\pm0.017$ and $-5.232\pm0.020$. Their
  measurements come from integrating the resolved spectra of the two
  components from optical to mid-IR
  wavelengths.\footnote{These values agree well with
    the \Lbol\ derived from resolved $K$-band photometry and our
    bolometric corrections in \S~\ref{sec:lbol}. Using the high quality
    parallax to \eInd~A
 ($\pi=276.1\pm0.3$~mas; \citealp{2007A&A...474..653V}), the
  $K_{MKO}$-band absolute magnitudes for the two components are $M_K =
  13.58\pm0.02$ and $15.84\pm0.03$~mag,
and $K$-band bolometric corrections are $2.87\pm0.10$ and
$2.20\pm0.10$~mag (including the intrinsic scatter and the
$\pm$0.5~subclasses uncertainty in the spectral types). Together, these
give $\log(\Lbol/\Lsun) = -4.67\pm0.04$ and $-5.31\pm0.04$~dex for
\eInd~Ba and~Bb, respectively.}

The effective temperatures of the two components have been derived by
fitting model atmospheres by two groups.
(1)~\citet{2009ApJ...695..788K} analyzed resolved near-IR ($JHK$)
spectroscopy of the two components using \citet{2005astro.ph..9066B}
model atmospheres computed with $\pm$50~K spacing. They identified the
best 3~fitting models to the data. We take the mean and RMS of their
results, $\Teff_{,A} = 1275\pm25$~K and $\Teff_{,B}=900\pm25$~K. The
25~K uncertainty is likely underestimated, as discussed in
\S~\ref{sec:teff}, especially as the fit to component~Ba does not appear
as good as that for component~Bb.
(2)~\citet{2010A&A...510A..99K} obtained resolved
  0.6--5.0~\micron\ spectra of the two components and compared the full
  spectra and the 1.25~\micron\ \ion{K}{1} absorption doublet to
  BT-Settl model atmospheres. Their comparisons applied an absolute flux
  calibration to the model atmospheres, based on the known luminosity
  and distance to the system. They found good agreement for
  $\Teff = $1300--1340~K and 880--940~K for the two components. We adopt
  the mean values and use the quoted range as the RMS.

Using the \citet{2010A&A...510A..99K} temperatures,
  Figure~\ref{fig:hrd-both} shows the position of the two components of
  \eInd~Bab on the H-R diagram, and Figure~\ref{fig:hrd-agediff-both}
  shows the age difference between the two components, computed in the
  same fashion as \S~\ref{sec:isochrone-2m1209}. Like for \twomassbinAB,
  the data are consistent with coevality for the binary, with a
  $\log(age)$ difference of $0.5_{-0.3}^{+0.4}$ and
  $0.3_{-0.4}^{+0.3}$~dex from the Lyon and Tucson models, respectively,
  though only at the 2$\sigma$ level when using the Lyon/COND models.
  The uncertainties are smaller for \eIndBab\ compared to \twomassbinAB\
  due to the much more precise distance and temperatures. The
  \citet{2009ApJ...695..788K} temperatures give a similar result for the
  component ages, in fact slightly more consistent with coevality,
  though not significantly so given the uncertainties.

Our conclusion of coevality for the system agrees with the (somewhat
different) approach of \citet{2009ApJ...695..788K}, who determine \logg\
and \Teff\ by fitting model atmospheres to the resolved spectra and then
use the results to compute the ages from evolutionary models.
However, note that \citet{2003ApJ...599L.107S} determine a hotter \Teff\
of $1500\pm100$~K for \eInd~Ba based on high resolution near-IR
spectroscopy; if this temperature is adopted, then this component's H-R
diagram position is off the model loci, with an age older than 10~Gyr
and inconsistent with coevality.

The age we derive from the H-R diagram is consistent with independent
constraints on the age from the primary star \eInd~A.
The most recently used estimate is 0.8--2.0~Gyr from
\citet{1999A&A...348..897L}, based on their semi-empirical relation
between stellar rotational period and age. Appendix~\ref{sec:epsInd-age}
updates the age estimate, leading to a broader range of 0.5--7.0~Gyr.
Both values for \eInd~A agree well with the median ages of 
  0.6--2.5~Gyr
derived by us from the H-R diagram analysis.

\subsection{Systematic Uncertainties in \Teff's for T Dwarfs from
    Model Atmospheres \label{sec:teff}}

For both \twomassbinAB\ and \eInd~Bab, while the models do successfully
indicate that the systems are coeval, the mass information derived from
the H-R diagram reveals some interesting issues. Table~\ref{table:ages}
gives the computed component masses and mass ratios, based on the same
calculations used to determine the ages.
  
At face value, the HR diagram-derived mass ratio for \twomassbinAB\
appears to be inverted, where the much brighter component~A is
non-sensibly indicated to be the lower-mass component. However, the
formal uncertainties are very large and at the $\approx1\sigma$ level,
consistent with the $q\approx0.5$ value inferred from evolutionary
models (\S~\ref{sec:properties} and Table~\ref{table:evolmodels}).
For \eInd~Bab, the mass ratio from the H-R diagram is
  in good agreement with the value of $q=0.60\pm0.02$ derived by
  \citet{2004A&A...413.1029M}, based on the component luminosities,
  evolutionary models, and an assumed age of 0.8--2.0~Gyr (where we have
  taken the average and RMS of their computed values).
So within the accuracy afforded by the current data, the mass ratios
derived from the H-R diagram seem reasonable.

However, a significant disagreement occurs when considering the total
mass of \eIndBab. The total mass derived from the H-R diagram is
$72_{-9}^{+10}$~\Mjup\ ($78_{-10}^{+11}$~\Mjup) as derived from the Lyon
(Tucson) models and the \citet{2009ApJ...695..788K} temperatures,
with the \citet{2010A&A...510A..99K} temperatures
  giving a slightly higher total mass of $86_{-10}^{+11}$~\Mjup\
  ($95_{-12}^{+13}$~\Mjup). (Note that the calculation of the total
mass from the H-R diagram is {\em independent} of the aforementioned
uncertainty in \eInd~A's age). Such values strongly conflict with the
preliminary dynamical mass of $121\pm1$~\Mjup\ measured by
\citet{2009AIPC.1094..509C} from the binary's relative orbital motion.
And thus, there is a problem in one or more of the input data or models
involved in the analysis. The observations of the component \Lbol's and
total dynamical mass seem secure. The \Lbol's are
  determined by \citet{2010A&A...510A..99K} using resolved spectroscopy
  across a very broad wavelength range, agree well with the values
  obtained using near-IR photometry with our well-determined bolometric
  corrections (\S~\ref{sec:isochrone-epsInd}), and the uncertainties
are fully propagated through our analysis. Likewise, the Cardoso \etal\
dynamical mass also seems reasonable given the large angular separation
between the two components, the large number of observing epochs, and
the long duration of the monitoring ($\approx$40\% of the derived
orbital period, which should be sufficient to determine an accurate
mass, e.g., as demonstrated by \citealp{2004A&A...423..341B} and
\citealp{2009arXiv0909.4784D}).

\begin{figure}
\centerline{\includegraphics[width=4.3in]{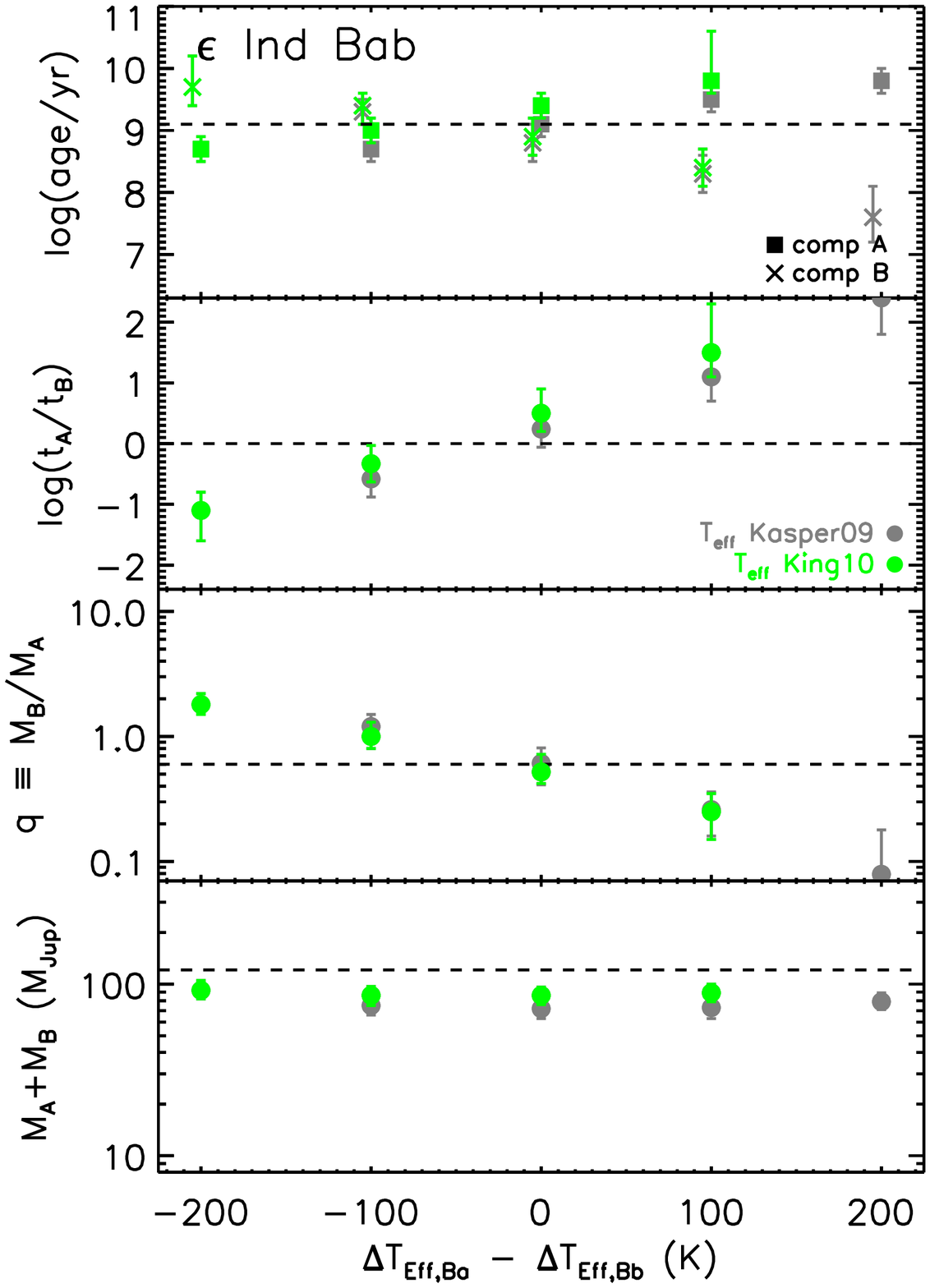}}
\vskip 2ex
\caption{\normalsize The effect of changing the relative \Teff\ values
  assigned to \eInd~Ba and~Bb on the physical properties derived from
  the H-R diagram positions, based on the Lyon/COND evolutionary models.
  The relative \Teff\ difference between the two components is assumed
  to be applied with an \underline{equal but opposite} amplitude to both
  components. For instance, the leftmost point represents a 50~K
  decrease in \Teff\ for component~Ba and a 50~K increase for
  component~Bb. The four plotted properties are the ages of the
  individual components, the age difference, the mass ratio, and the
  total mass. The horizonal dashed lines represent the available
  constraints: for this system, the age of the primary star
  ($\log(t/{\rm yr})=9.1$), the requirement of coevality
  ($\log(t_A/t_B)=1$), the mass ratio inferred from the luminosity ratio
  ($q=0.6$), and the total dynamical mass (121~\Mjup).
  The grey points are the results based on using the
    \citet{2009ApJ...695..788K} \Teff's and the green on the
    \citet{2010A&A...510A..99K} \Teff's. The rightmost point is missing
    for the \citet{2010A&A...510A..99K} models as such a large
    temperature change moves the binary components off the loci of
    available models.
  \label{fig:plot-teff-changes-epsInd-Trelative}}
\end{figure}

\begin{figure}
\centerline{\includegraphics[width=4.3in]{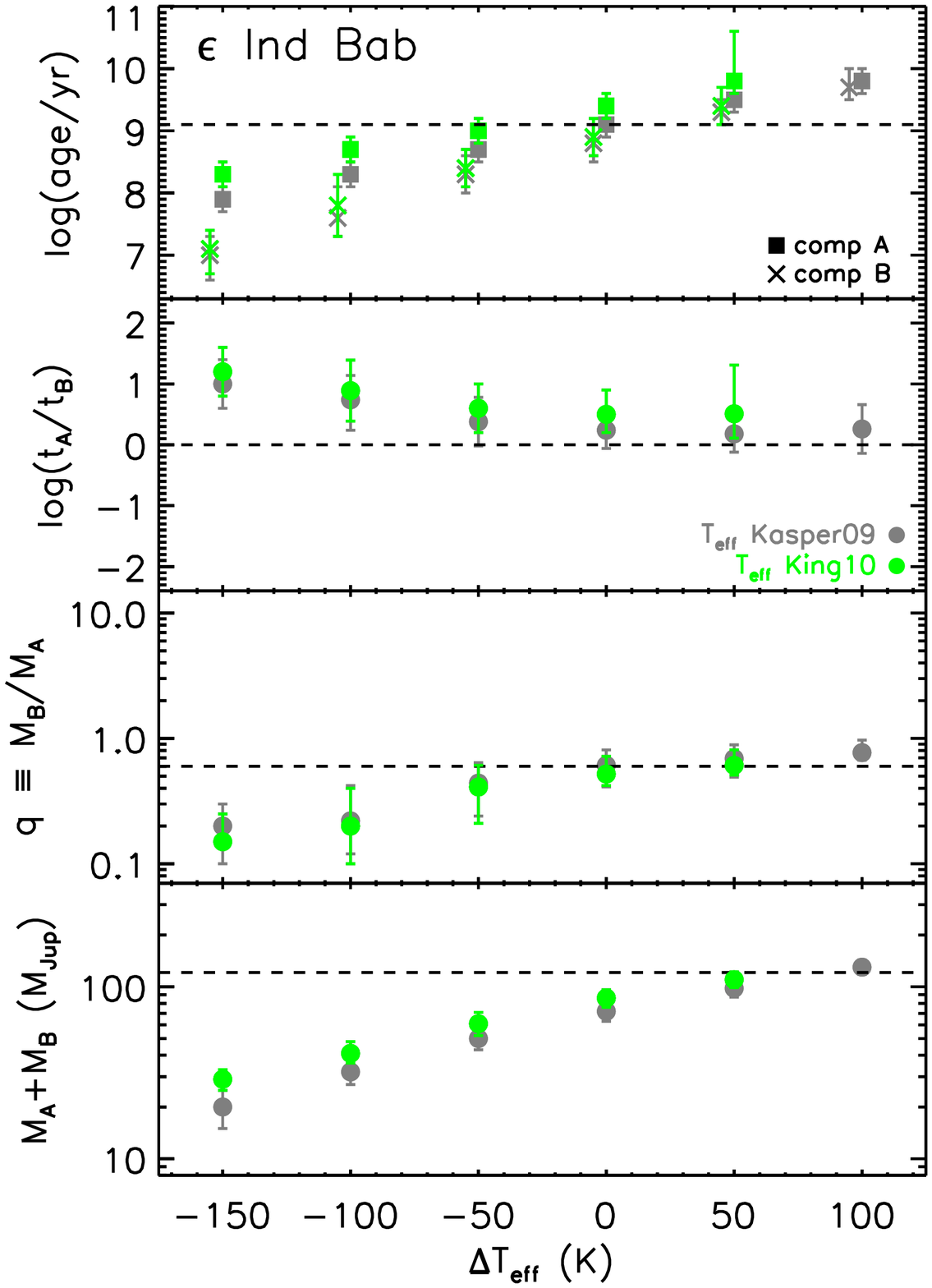}}
\vskip 2ex
\caption{\normalsize The effect of applying a constant offset to the
  assumed \Teff\ values for \eInd~A and \eInd~B, with
    grey being \Teff\ from \citet{2009ApJ...695..788K} and green from
    \citet{2010A&A...510A..99K}. For example, the leftmost point
  represents a 150~K decrease from the nominal \Teff\ values employed in
  \S~\ref{sec:isochrone-2m1209}. See
  Figure~\ref{fig:plot-teff-changes-epsInd-Trelative} caption for more
  details.
  \label{fig:plot-teff-changes-epsInd-Tshift}}
\end{figure}

\begin{figure}
\hskip -0.1in
\includegraphics[width=3.2in]{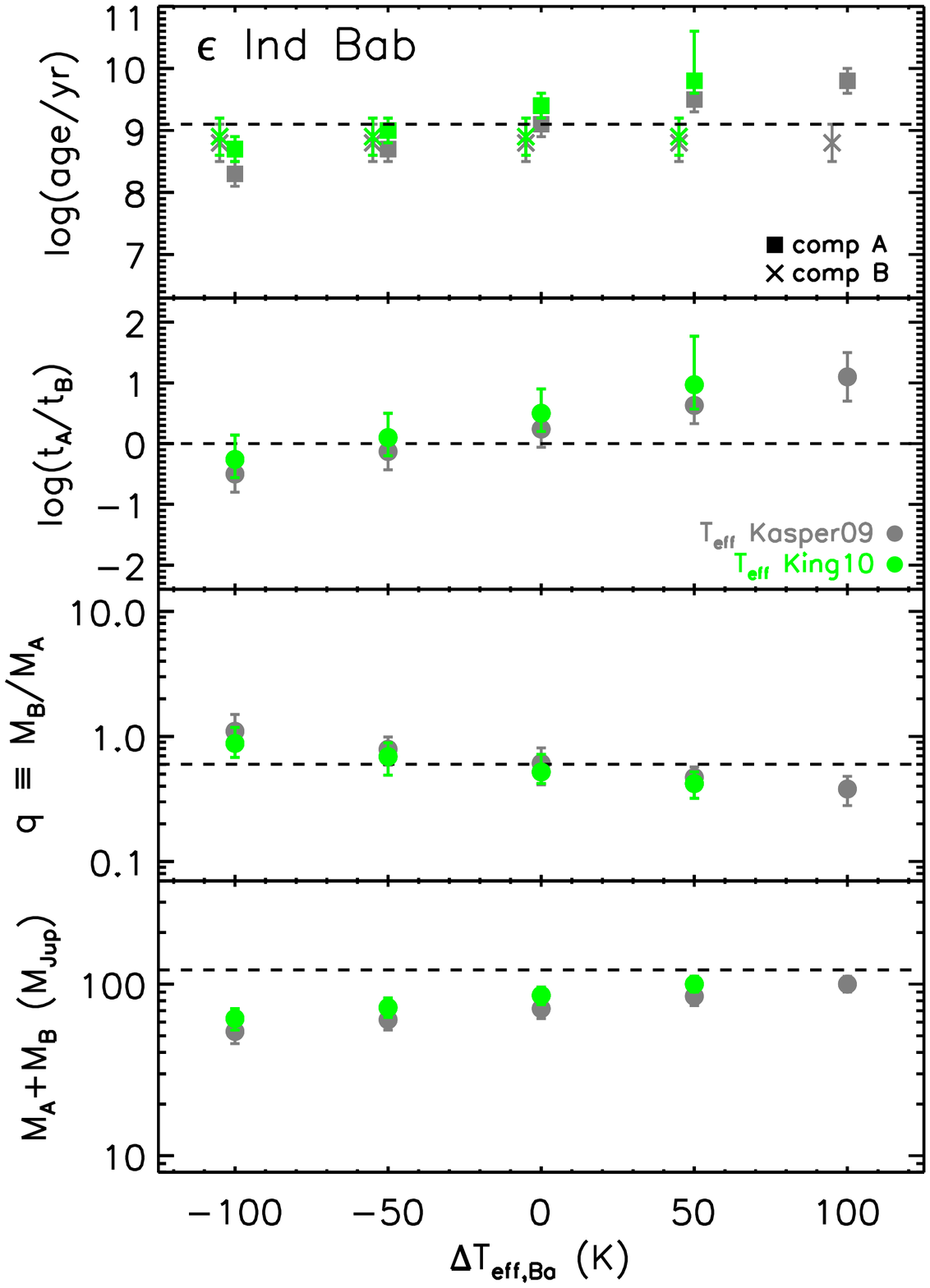}
\hskip 0.3in
\includegraphics[width=3.2in]{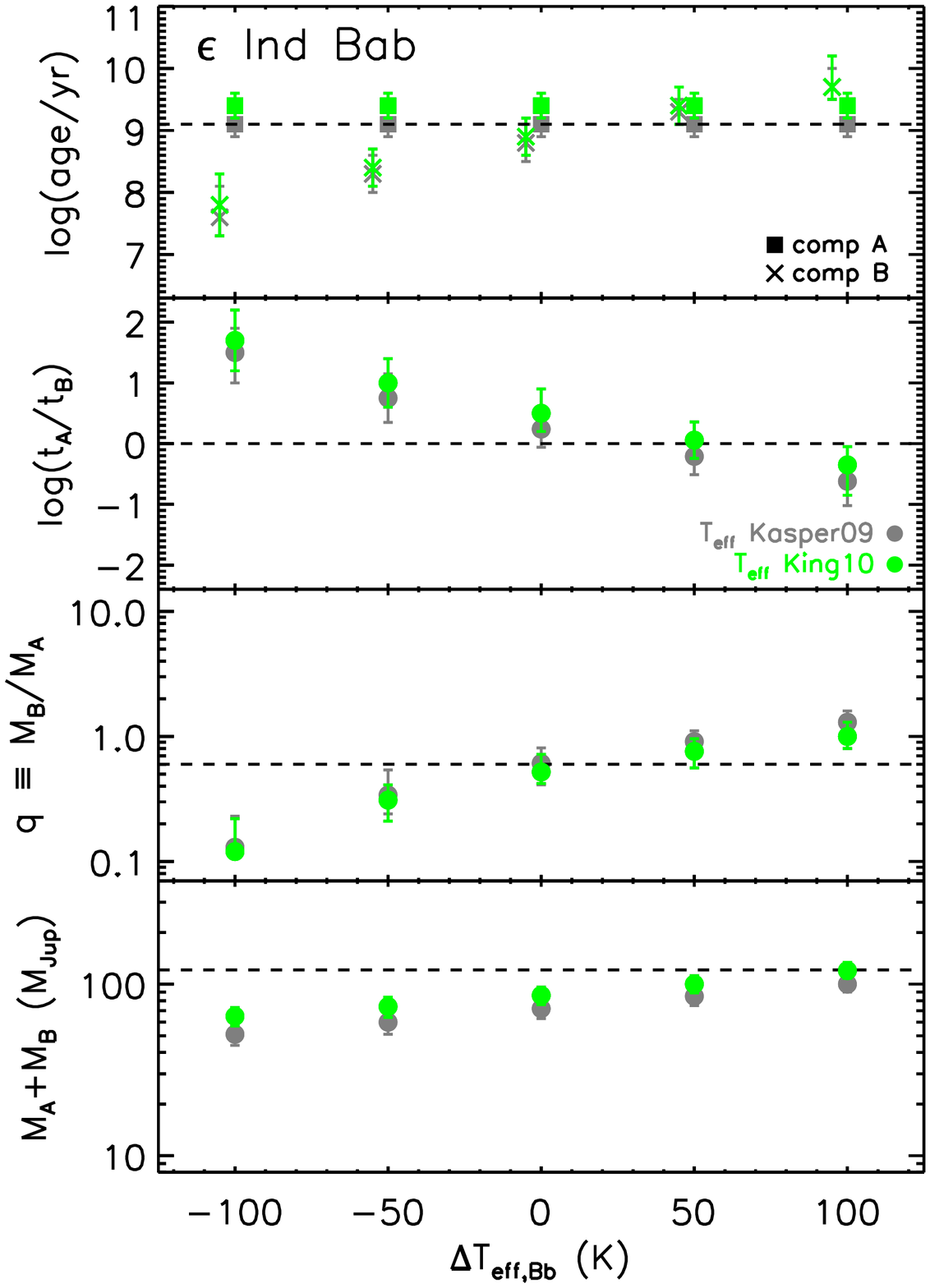}
\vskip 2ex
\caption{\normalsize The effect of changing the assumed \Teff\ value for
  one component of \eIndBab\ while leaving the \Teff\ of the other
  component fixed, with grey points from the
    \citet{2009ApJ...695..788K} values and green from
    \citet{2010A&A...510A..99K}. {\bf Left:} Changing the \Teff\ of
  component~Ba. For example, the leftmost point represents a 100~K
  decrease from the nominal \Teff\ value employed in
  \S~\ref{sec:isochrone-2m1209}. See
  Figure~\ref{fig:plot-teff-changes-epsInd-Trelative} caption for more
  details. {\bf Right:} Changing the \Teff\ of component~Bb.
  \label{fig:plot-teff-changes-epsInd-both}}
\end{figure}

Assuming for the moment that the loci of evolutionary models are
correct, this discrepancy in total mass of \eIndBab\ must arise from
errors in the temperatures assigned to the binary components from the
model atmosphere fitting.\footnote{A more exotic explanation would be
  that \eIndBab\ is an unresolved triple system, with one of the
  components being an unresolved, nearly equal-mass binary. This would
  increase the total mass of the system derived from the H-R diagram,
  since correcting for such binarity would reduce the \Lbol\ of the
  component and lead to a higher mass for a fixed \Teff. However, such a
  change would drive the H-R diagram positions towards non-coevality
  (assuming the assigned \Teff's are correct). More significantly, the
  resolved absolute IR magnitudes of the two known components of
  \eIndBab\ are consistent with other T~dwarfs of comparable spectral
  types (\eg, Figure~16 of \citealp{2006ApJS..166..585B}), and thus
  correction for any unresolved binarity would make the associated
  component underluminous.} Generally speaking, changes to the relative
temperatures between the two components would alter the inferred mass
ratio, whereas systematic shifts to the temperatures assigned to both
components would alter the derived total mass.
Figures~\ref{fig:plot-teff-changes-epsInd-Trelative}
and~\ref{fig:plot-teff-changes-epsInd-Tshift} summarize the changes to
the HR-diagram-derived properties of the \eIndBab\ system (component
ages, age difference, mass ratio, and total mass) if the \Teff's for the
two components are changed in these three ways: relative changes in the
\Teff's, overall shifts in the \Teff's, and changes to
    the \Teff\ of only one component. These calculations place the
  following constraints on the \Teff\ estimates:

\begin{itemize}
  
\item The {\em relative} \Teff\ difference between
    \eInd~Ba and~Bb appears to be consistent with most of the
    constraints, namely coevality, an estimated age of $\approx$1~Gyr,
    and a mass ratio of $\approx$0.6
    (Figure~\ref{fig:plot-teff-changes-epsInd-Trelative}). However, the
    total mass of the system is under-predicted by the evolutionary
    models using these \Teff's.
  
\item The disagreement between the total mass from the
    H-R~diagram and the orbital monitoring disappears if both components
    are $\approx$50--100~K hotter
    (Figure~\ref{fig:plot-teff-changes-epsInd-Tshift}), which still
    leaves the system coeval. However, an older age of $\approx$6~Gyr
    then results.
  
\item Changing the \Teff\ for only one of the
    components cannot simultaneously fulfill the requirements of
    coevality and agreeing with the dynamically measured total mass
    (Figure~\ref{fig:plot-teff-changes-epsInd-both}).
  
\end{itemize}

Altogether, this analysis points to a small, but non-zero, systematic
error of $\approx$50--100~K in the published \Teff\ values from model
atmosphere fitting.
This seems plausible given the known inadequacies of the model
atmospheres, especially in the L/T transition region
(\S~\ref{sec:isochrone}), and thus invoking errors in the evolutionary
models is not compelling. The implied temperatures for the two
components would then be $\approx$1350--1400~K and $\approx$950-1000~K
for components~Ba and~Bb, respectively.
For component~Ba, this hotter temperature is consistent with the
$1500\pm100$~K found by \citet{2003ApJ...599L.107S} and noticeably
hotter than previous fitting of early-T~dwarfs ($\approx$1100--1200~K;
\S~\ref{sec:isochrone-2m1209}).
For component~Bb, the implied hotter temperature agrees with that found
expected for mid/late-T dwarfs based on model atmosphere fitting by
\citet{2006ApJ...639.1095B}, where a linear fit to their results gives
1030~K for the T6~subclass (Equation~7 of \citealp{liu08-2m1534orbit}).

The systematic \Teff\ errors that we infer from the isochrone method are
comparable to the estimated uncertainties in spectral fitting of field
T~dwarfs (\S~\ref{sec:isochrone-2m1209} and references therein), though
\eIndBab\ provides a more stringent test of the models given this
binary's abundance of observational constraints compared to ordinary
field brown dwarfs.\footnote{A somewhat similar, though less precise,
  consistency check of the atmospheric model fitting relative to the
  evolutionary models comes from the derived ages of field and benchmark
  T~dwarfs, where the fitted \Teff\ and \logg\ values combined with
  evolutionary models lead to reasonable age determinations of a few to
  several~Gyr for single field objects
  \citep[e.g.][]{2008arXiv0804.1386L, 2007arXiv0705.2602L,
    2006astro.ph.11062S, 2006ApJ...639.1095B}.} Also, note that the
$x$-axes in
Figures~\ref{fig:plot-teff-changes-epsInd-Trelative}--~\ref{fig:plot-teff-changes-epsInd-both}
show the full possible range of \Teff\ changes; larger changes would
displace one or both binary components off the model loci on the H-R
diagram, leading to ages that are either far too young
($\lesssim$10~Myr) or too old ($\gtrsim$10~Gyr).
%


Finally, we point out that given the measured \Lbol's and dynamical mass
for \eIndBab, Figure~\ref{fig:teff-logg} shows that the evolutionary
models would lead to an inferred age of about 5~Gyr. In the same
fashion, King et al. (2010) have inferred an age of 3.7-4.3 Gyr from the
evolutionary models (see also Cardoso et al. 2009). This is at the older
end of the age range inferred for the K5V primary star \eInd~A
(Appendix~\ref{sec:epsInd-age}), though the age-dating methods are
largely derived for $\lesssim$1~Gyr solar-type stars and are not
well-calibrated for such old ages or for such a late-type primary. In
fact, it may be that the age of the \eInd~ABab system is most accurately
determined from brown dwarf evolutionary models, though it will be
difficult to test this hypothesis independently.


\section{Conclusions}

We have identified the T3~dwarf \twomassbin\ as a tight substellar
binary, with an extreme mass ratio ($q\approx0.5$) compared to
previously known field brown dwarf binaries. The near-IR photometry
combined with modeling of the integrated-light spectrum indicates
spectral types of T2.0$\pm$0.5 and T7.5$\pm$0.5 for the two components.
This newly discovered binarity lessens the utility of this object as the
primary near-IR spectral-type standard for T3~dwarfs. We suggest
SDSS~J1206+2813 as a replacement.

The highly unequal mass ratio of the system allows us to test the joint
accuracy of evolutionary and atmospheric models, by examining if the
ages inferred from the H-R diagram position of the two components are
consistent with coevality. Using the photometric distance to the system
and latest available \Teff's from model atmosphere fitting of T~dwarfs,
we find that the models successfully indicate that the two components of
\twomassbinAB\ are coeval, with a difference of $\log(age) = -0.8\pm1.3$
($-1.0_{-1.3}^{+1.2}$)~dex based on the Lyon (Tucson) models. A similar
analysis of the T1+T6 binary \eIndBab\ also supports the accuracy of the
models, with the system being consistent with coevality
  at the 1--2$\sigma$ level, depending on the choice of model atmosphere
  fits and evolutionary models. (The temperatures for \eInd~Bab from
  \citealp{2010A&A...510A..99K} in combination with the Lyon/COND models
  produce the most non-coeval results, with an age difference of
  $\log(age)=0.5^{+0.4}_{-0.3}$~dex.)

While the models succeed in respect to the isochrones, for \eInd~Bab the
total mass derived from the H-R diagram ($\approx$80~\Mjup) is
discrepant with the measured dynamical mass of $121\pm1$~\Mjup.
This is most simply resolved by assuming a systematic
  increase in \Teff\ for the two components by $\approx$50--100~K
  warmer, \ie, an increase to $\approx$1375~K and 950~K for components~Ba
  and~Bb, respectively. Such an error is not unexpected, given the known
inadequacies of the model atmospheres in the L/T transition region, the
quality of the spectral fit, and the general uncertainties in the
procedures used to fit such models to observed spectra. Such a \Teff\
increase leads to an implied age of about 6~Gyr for the system, somewhat
older than previous estimates.

While the $\approx$50--100~K errors are comparable to those typically
cited in the atmospheric model fitting, it is worth emphasizing that
\eIndBab\ indicates that such errors are present (\ie. non-zero), as the
masses inferred from the H-R diagram are strongly discrepant with the
dynamical mass, {\em a result that is independent of the uncertainty in
  the age of the system.}
However, assuming the evolutionary models are basically correct, \Teff\
errors much larger than $\approx$100~K are not warranted, as the
inferred properties from the H-R diagram would then be inconsistent with
the underlying constraints of coevality and a mass ratio $\le$1.0.
Also note that the very precise \Teff\ results from the
  T5+T5.5 mass-benchmark binary 2MASS~J1534$-$2952AB and the M4+T8.5
  age-benchmark system Wolf~940AB point to $\approx$100~K overestimates
  of \Teff\ when model atmospheres are used to fit the near-IR spectrum
  of T~dwarfs \citep{liu08-2m1534orbit, burningham08-T8.5-benchmark},
  similar in amplitude though opposite in sign to what we find here from
  \eIndBab.
For Wolf~940B, \citet{2010arXiv1007.1252L} show that
luminosity-constrained model fits to the near-IR spectrum of this T8.5
dwarf are poor, with the synthetic spectrum $J$ and $H$-band peaks too
high and the $K$-band peak too low. Most likely this is due to the known
incompleteness of the molecular opacity line lists in this region.

Thus, substantial errors in evolutionary and/or atmospheric models for
T~dwarfs do not appear to be required, though it is not
  possible to fully reconcile all the model predictions with the
  observations, especially for \eInd~Bab where the measurements are
  exceptionally precise.
It is also worth noting that \citet{2008arXiv0807.2450D} have found that
the luminosities predicted by evolutionary models at $\approx$0.8~Gyr
may be systematically underestimated by a factor of 2--3, based on the
luminosities and dynamical masses of the L4+L4 substellar binary
HD~130948BC. A simple factor of~3 boost to the model-predicted
luminosities seems implausible --- the model tracks on the H-R diagrams
(Figure~\ref{fig:hrd-both}) would then give implausibly old ages
($>>$10~Gyr). Such a luminosity error could be made compatible with the
H-R diagram positions if the current effective temperatures for the
T~dwarfs were also systematically overestimated by $\gtrsim$200~K, but
such a large discrepancy seems implausible.

Even stronger constraints on the models from the isochrone test will
require smaller uncertainties in \Lbol\ and \Teff. For \twomassbinAB,
numerical tests using the measurements from
\S~\ref{sec:isochrone-2m1209} indicate that a $\approx$5\% distance and
$\approx$25~K uncertainty in \Teff\ are needed to measure a 2$\sigma$
discrepancy from coevality. In absence of a parallax, our analysis is
consistent with coevality, but the results on absolute ages in
Table~\ref{table:evolmodels} cannot be taken as a firm constraint. The
parallax measurement is within current capabilities
\citep[e.g.][]{2004AJ....127.2948V} and is being
  pursued by us using the Canada-France-Hawaii Telescope
  \citep[e.g.][]{2009AAS...21340617D}, but the required \Teff\ accuracy
is a challenge to atmospheric models. Both \twomassbinAB\ and \eIndBab\
have plausibly short orbital periods for dynamical mass determinations,
which will significantly strengthen the future tests of the models. This
is especially true in the case of \eInd~Bab, given the additional
independent constraint from the age estimate of the primary star.

The immediate prospects of finding similarly extreme systems to further
test the models in this fashion are limited, given that hundreds of
nearby ultracool dwarfs have already been imaged at high angular
resolution but yielding only \twomassbinAB\ and \eInd~Bab. As one
alternative path, \citet{2008arXiv0803.0295B} have pursued the
possibility of finding previously unrecognized extreme mass-ratio
ultracool binaries by modeling of integrated-light spectra, suitable for
special cases where the primary and secondary spectral types are
sufficiently discrepant (though the resolved measurements of the
individual components needed to place them on the H-R diagram may be
challenging). In the near-future, the advent of powerful new wide-field
surveys, such as the Pan-STARRS optical survey
\citep{2002SPIE.4836..154K} and the WISE mid-IR survey satellite
\citep{2009AAS...21345907M}, will boost the prospects for expanding the
census of field brown dwarfs and thereby enable discovery of new
binaries. A larger sample of extreme systems will allow the isochrone
test to be applied over a wider range of temperatures and luminosities,
thus challenging the models over a large span of brown dwarf masses and
ages.


\acknowledgments

We gratefully acknowledge the Keck LGS AO team for their exceptional
efforts in bringing the LGS AO system to fruition. It is a pleasure to
thank Jim Lyke, Hien Tran, Heather Hershley, Gary Punawai, Al Conrad,
Randy Campbell, Jason McIlroy, and the Keck Observatory staff for
assistance with the observations. This work has benefited from
discussions with Michael Cushing, Markus Kasper, and Jon Swift. We thank
David Pinfield, Nicolas Lodieu, Phillipe Delorme, Ben Burningham, and
Steve Warren for providing electronic version of published late-T dwarf
spectra.
Our research has employed the 2MASS data products; NASA's
Astrophysical Data System; the SIMBAD database operated at CDS,
Strasbourg, France; the M, L, and T~dwarf compendium housed at
DwarfArchives.org and maintained by Chris Gelino, Davy Kirkpatrick,
and Adam Burgasser \citep{2003IAUS..211..189K, 2004AAS...205.1113G};
the SpeX Prism Spectral Libraries maintained by Adam Burgasser at
http://www.browndwarfs.org/spexprism; and the VLM Binaries Archive
maintained by Nick Siegler at http://www.vlmbinaries.org.
MCL and TJD acknowledge support for this work from NSF grant
AST-0507833 and an Alfred P. Sloan Research Fellowship.  
SKL's research is supported by the Gemini Observatory, which is operated
by the Association of Universities for Research in Astronomy, Inc., on
behalf of the international Gemini partnership of Argentina, Australia,
Brazil, Canada, Chile, the United Kingdom, and the United States of
America.
Finally, the authors wish to recognize and acknowledge the very
significant cultural role and reverence that the summit of Mauna Kea has
always had within the indigenous Hawaiian community.  We are most
fortunate to have the opportunity to conduct observations from this
mountain.

{\it Facilities:} Keck II Telescope (LGS AO, NIRC2)




\appendix

\section{Updated Near-IR Bolometric Corrections for Ultracool Dwarfs \label{sec:bolcorr}} 

\begin{figure}
\vskip -1in
\centerline{\includegraphics[width=4.5in,angle=90]{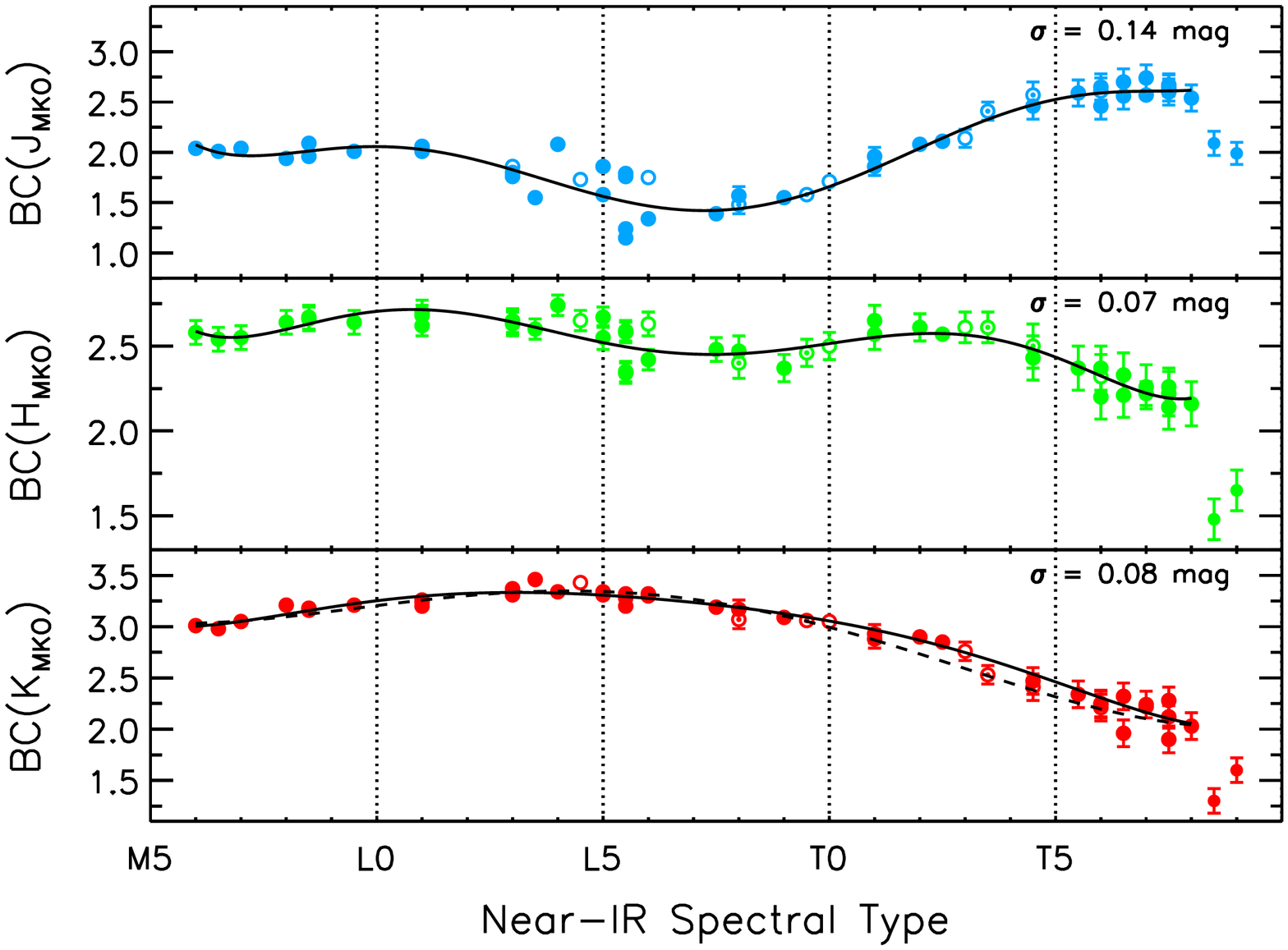}}
\vskip 2ex
\caption{\normalsize Bolometric correction as a function of near-IR
  spectral type for the MKO $J$, $H$, and $K$-band filters, based on
  data from \citet{gol04}, \citet{2006astro.ph..9464L},
  \citet{2006liu-hd3651b}, \citet{2008arXiv0804.1386L},
  \citet{2010A&A...511A..30S}, and \citet{2010arXiv1007.1252L}. Points
  without visible error bars have uncertainties smaller than the
  plotting symbol. The solid line overplots a 6th-order polynomial fit
  to the datasets for objects as cool as T8, with coefficients given in
  Table~\ref{table:bolcorr}. The dotted line for the $K$-band data gives
  the fit from \citet{gol04}. Known and suspected binaries are plotted
  as open circles and as circumscribed small circles, respectively, and
  excluded from the fits. Suspected binaries are those flagged by
  \citet{2006astro.ph..5037L} based on their large overluminosity
  relative to objects of similar spectral type. The L~dwarfs are
  classified on the \citet{geb01} scheme and the T~dwarfs on the
  \citet{2005astro.ph.10090B} scheme. The two coolest
    objects (T8.5 Wolf~940B and T9 ULAS~0034$-$0052) are excluded from
    the fit.\label{fig:bolcorr}}
\end{figure}

We improve the polynomial fit for the $K$-band bolometric correction as
a function of near-IR spectral type from \citet{gol04}, by updating the
spectral types for the T~dwarfs in their sample to types from
\citet{2005astro.ph.10090B}; adding data for the T~dwarfs HD~3651B and
HN~Peg~B \citep{2006liu-hd3651b, 2006astro.ph..9464L,
  2008arXiv0804.1386L}; and constructing fits for the $J$ and $H$-band
bolometric corrections.
We use a 6th-order polynomial fit to accurately capture the behavior of
the $J$ and $H$-band data and to avoid artificially steep changes in the
polynomials at the earliest and latest spectral types. The results of
the new fits are shown in Figure~\ref{fig:bolcorr} and
Table~\ref{table:bolcorr}. Note that the use of the revised spectral
types for the T~dwarfs leads to 30--50\% smaller RMS about the fit
compared to the original Golimowski \etal\ study.

We also plot the $BC$ values for two of the latest type
  objects known, the T8.5~dwarf Wolf~940B
  \citep{burningham08-T8.5-benchmark, 2010arXiv1007.1252L} and the
  T9~dwarf ULAS~0034$-$0052 \citep{2007MNRAS.381.1400W,
    2010A&A...511A..30S}. From the published bolometric luminosities,
  MKO photometry, and distances, we compute their bolometric corrections
  in $JHK$, though these are not included in the polynomial fit. The
  values are notably different from even the T7--T8 objects, likely due
  to the sharply increasing fraction of \Lbol\ emitted at mid-infrared
  wavelengths in the very latest known objects (e.g. Figure~1 of
  \citealp{2010ApJ...710.1627L}) and perhaps improvements in the methods
  used to derive \Lbol\ for late-T dwarfs using model atmospheres since
  the \citet{gol04} study, which assumed a Rayleigh-Jeans approximation
  beyond 5~\micron.


\section{Homogenous Compilation of Substellar Field
    Binaries \label{sec:binaries}}

We determined or compiled from the literature the spectral types,
distances, and luminosities of all ultracool field binaries with
integrated-light spectral types of L4 or later or those with ancilliary
evidence for being substellar. The available literature compilations are
based on a broad mix of input data and calculation methods.
To improve upon this, we assembled a more homogenous compilation of
binary properties, as described below. The final results are in
Table~\ref{table:binaries}.

\vskip 1ex
\noindent{\bf Photometry:} For the integrated-light photometry, we
preferentially used measurements on the MKO photometric system where
available and 2MASS measurements for the remaining objects.

To determine the resolved photometry, we compiled the highest precision
$JHK$ flux ratio measurements available from the literature and
supplemented them with measurements from public data archives (\HST,
VLT, and Gemini) and our own unpublished Keck LGS AO data. In cases
where the integrated-light photometry and flux ratios were not on the
same photometric system, we used conversions between MKO and 2MASS from
\citet{2004PASP..116....9S}.

Infrared data were not available for 5~binaries originally discovered by
\HST\ using the $F814W$ bandpass \citep[e.g.][]{2001AJ....121..489R,
  2003AJ....125.3302G, 2003AJ....126.1526B}. For these cases, we
estimated the $K$-band flux ratio from the optical results. To do this,
we assumed $\Delta$F814W$ = \Delta{I}$, which is accurate to
$\approx0.02$~mag according to \citet{2003AJ....126.1526B}. To estimate
the secondary's spectral type, we derived the relation between spectral
type and $M_I$, based on the 11~objects that were used in the Liu et al.
(2006) faint relations and have $I$-band photometry in
\citet{2002AJ....124.1170D}: \begin{eqnarray}
  M_I & = & 14.83~+~(6.396\times10^{-1})\times{\rm SpT}~-~(1.542\times10^{-2})\times{\rm SpT}^2
\end{eqnarray}
where $SpT=0$ means L0, etc, using optical spectral types for the
L~dwarfs. The fit is valid from L0 to T7.5, and the rms about the fit is
0.25~mag.
Then using the adopted primary spectral type (described below) and the
observed $\Delta{I}$, we derived the secondary spectral type. With
spectral types for both components, we use the
\citet{2006astro.ph..5037L} SpT--absolute magnitude relations to
estimate $\Delta{K}$ for the binary. We accounted for the additional
uncertainty introduced by this process in our analysis, and the affected
objects are marked in Table~\ref{table:binaries} as being
correspondingly uncertain.

\vskip 1ex
\noindent{\bf Resolved spectral types and distances:}
When available, we adopt the individual component spectral types
previously determined from detailed analysis, e.g., spectral
decomposition.

For the remaining binaries, we assumed that the primary spectral types
and uncertainties were identical to the integrated-light spectral types.
We adopted optical spectral types determined on the
\citet{1999ApJ...519..802K} classification scheme for L~dwarfs and
infrared types on the \citet{2005astro.ph.10090B} classification scheme
for T~dwarfs, except in two cases.
One was DENIS-P~J225210.7$-$173013AB, which \citet{2004A&A...416L..17K}
classified as L7.5 based on a comparison of the $H$ and $K$-band spectra
to those of optically typed L~dwarfs.
The other was 2MASS~J09201223$+$3517429AB, which has highly discrepant
integrated-light spectral types of L6.5 in the optical
\citep{2000AJ....120..447K} and T0p in the infrared. We adopted the
infrared type in this case.

To derive the secondary spectral types, we used the 
  average of the ``faint'' and ``bright'' SpT--absolute magnitude
relations of \citet{2006astro.ph..5037L} along with the known or
estimated absolute magnitudes for the secondaries.  (The
  Liu \etal\ relations were extended to objects of M6--L1 using
  photometry from the \citealp{2010ApJ...710.1627L} literature
  compilation.)
For binaries with parallax measurements, we computed the absolute
magnitudes of the secondary components directly. For the rest, we
estimated photometric distances based on the primary components'
apparent magnitudes and the same averaged \citet{2006astro.ph..5037L}
SpT--absolute magnitude relations. If distance estimates were available
from multiple bandpasses, we preferred the $K$-band estimate, then
$H$-band, and finally $J$-band, to minimize the potential non-monotonic
behavior in the L/T transition region. The uncertainties in the
resulting distances account for the input photometry uncertainties and
the intrinsic scatter in the empirical relations.

Using the final assembled IR absolute magnitudes for the secondaries, we
estimated spectral types using the method described in Section~3.2 of
\citet{2009ApJ...699..168D}. For a given bandpass, we computed the
spectral type probability distribution for an object based on its
absolute magnitude, the uncertainties in both the photometry and the
distances, and the intrinsic scatter in the empirical SpT-absolute
magnitude relations. We then combined the results from all available
bandpasses to form the joint spectral type probability distribution and
thus to estimate the final secondary spectral types and uncertainties.
We rounded all spectral types and their uncertainties to the nearest
0.5~subclass.

\vskip 1ex
\noindent {\bf Bolometric luminosities and luminosity ratios:}
We derived individual luminosities for the binary components from each
available bandpass ($JHK$) using the resolved photometry, distances, and
bolometric corrections. Bolometric corrections were computed based on
the polynomial fits in Appendix~\ref{sec:bolcorr} and the spectral types
assigned to each component. For computing luminosities, we assume
$M_{{\rm
    bol}, \odot} = 4.75$~mag. Uncertainties for the resulting \Lbol's
were computed from the uncertainties in the resolved photometry, in the
bolometric corrections due to the spectral type errors, and the
intrinsic scatter in the bolometric correction relations.

We also computed luminosity ratios for each available bandpass; this is
typically known to higher precision than \Lbol\ since the ratios are
independent of the distance uncertainty. The error in the luminosity
ratio comes from the quadrature sum of the errors in the flux ratios and
the bolometric corrections of the primaries and secondaries. For our
final adopted luminosities and luminosity ratios, we chose the bandpass
that gave the smallest error in the luminosity ratio.

\vskip 1ex
\noindent {\bf Mass ratios:} We used the \citet{1997ApJ...491..856B}
evolutionary models to derive the mass ratios for each system, using the
derived \Lbol's, their uncertainties, and an adopted age. 
  We take care to treat the distance uncertainty as an error common to
  each binary, as opposed to applying it independently to each
  individual component. For objects with specific age information in
the literature, we use that; otherwise we assume 1~Gyr. In both cases,
we ignore the age uncertainty in calculating the mass ratio,
since this is common to (most of) the objects. Four
binaries have their total mass measured dynamically, and we use their
published mass ratio measurements in these cases: 2MASS~J1534$-$2952AB
\citep{liu08-2m1534orbit}, HD~130948BC \citep{2008arXiv0807.2450D},
LP~349$-$25AB, and GJ~569Bab \citep{2010arXiv1007.4197D}.

To summarize the mass ratio results, we fit the observed values with a
power-law distribution.\footnote{We exclude the L6p+T7.5p system,
  SDSS~J141624.08$+$134826.7AB, from this ensemble analysis given its
  unusually large separation and mass ratio compared to all other field
  systems. Also, because of its large \Lbol\ ratio, its mass ratio
  estimate is highly dependent on the assumed age, from 0.2 to 0.5 for
  ages of 1~to~10~Gyr.} Since the sample is relatively sparse, instead
of the usual process of fitting to a binned histogram, we adopt a
maximum likelihood approach. The probability distribution of the mass
ratio $q$ is described by
\begin{equation}
P(q) = (\alpha +1)\ q^\alpha
\end{equation}
for $q=0-1$, with the constant chosen so that the total probability is
unity.  We define the log of the likelihood function as
\begin{eqnarray}
{\cal L} \equiv \ln P(q|\alpha) & = & \ln \prod_{i=1}^n (\alpha+1)\ q_i^\alpha \\
                                & = & \sum_{i=1}^{n} \Big[ \ln (\alpha+1) + \alpha \ln q_i \Big]  \\
                                & = & n \ln (\alpha+1) + \alpha \sum_{i=1}^{n} \ln q_i 
\end{eqnarray}
where $n$ is the number of objects in the sample and $q_i$ is the mass
ratio for each object. To find the best fitting $\alpha$, we find the
maximum of ${\cal L}$:
\begin{eqnarray}
\frac{d{\cal L}}{d\alpha} = \frac{n}{\alpha+1} + \sum_{i=1}^n \ln q_i = 0 
\end{eqnarray}
which gives the final result
\begin{eqnarray}
\alpha = -\Bigg[\ 1 + {n \over \sum\limits_{i=1}^n \ln q_i}\ \Bigg] .
\end{eqnarray}
For our sample, the distribution ${\cal L(\alpha)}$ is well-fit by a
gaussian, from which we extract 1$\sigma$ confidence limits. To
incorporate the uncertainties in $q_i$, we use a Monte Carlo approach,
computing the best $\alpha$ many times with the $q_i$ values drawn from
normal distributions; the resulting RMS dispersion in $\alpha$ from the
ensemble of results is much smaller than the 68\% confidence interval in
any realization of ${\cal L}$. As a simple approximation, we add the RMS
($\sigma_\alpha=0.3$) in quadrature to the confidence limits
($\sigma_\alpha=0.6$) to obtain our final uncertainty on $\alpha$.

We find $\alpha = 4.9 \pm 0.7$ for the sample in
Table~\ref{table:binaries}. For such an exponent, binaries with
$q\le0.65$ such as \eIndBab\ and \twomassbinAB\ amount to only 9\% of
the population.
Our result is comparable to, though somewhat steeper, than the
$q^{(4.2\pm1.0)}$ found by \citet{2006ApJS..166..585B} for a sample of
30~binaries with estimated primary masses of $<$0.075~\Msun. There are
some differences in the fitting method and the input data; for the
latter, the Burgasser \etal\ compilation included young binaries in open
clusters and star-forming regions (which tend to have a broader range of
mass ratios) and used the heterogenous set of mass ratios published in
the literature. Burgasser \etal\ also corrected the observed mass ratio
distribution to account for the bias towards equal-mass systems in
magnitude-limited surveys, based on an approximate conversion between
flux ratio and mass ratio \citep[see][]{2003ApJ...586..512B}. We choose
not to apply any corrections, since the surveys involved in
Table~\ref{table:binaries} span a wide range of wavelengths and
telescopes. Given these differences, a more shallow power-law is
expected from the Burgasser \etal\ compilation.
Similarly, using a full Bayesian analysis to account for differing
survey sensitivities, \citet{2007ApJ...668..492A} derived
$\alpha=1.8\pm0.6$; his sample included many late-M (stellar) primaries
which likely led to a shallower exponent.

\section{Summary of Age Estimates for \eInd~A \label{sec:epsInd-age}}

Age determinations for the K5V star \eInd~A are necessarily indirect and
rely on empirical correlations between stellar activity, rotation, and
age. \citet{1999A&A...348..897L} estimated a mean \hbox{log(age) =
  9.1} (1.3~Gyr) with a range of 8.9~--~9.3 (0.8~--~2.0~Gyr), based on
their semi-empirical relation between stellar rotational period and age
and the {\em estimated} rotational period of \eInd~A from
\citet{1997MNRAS.284..803S} as derived from \ion{Ca}{2} measurements and
an empirical relation between chromospheric activity and rotation. This
age range has been used for analysis of \eInd~Bab in previous work
\citep[e.g.][]{2009ApJ...695..788K,
  2004A&A...413.1029M, 2003A&A...398L..29S}.  

Additional age estimates come from the correlation between age and
chromospheric activity traced by \ion{Ca}{2} line emission, as measured
by the \Rhk\ index. \citet{1996AJ....111..439H} measured \eInd~A on
three occasions spanning 1.3~yr,
with a mean and an rms of $\log{\Rhk} = -4.56 \pm 0.01$~dex.
Chromospheric activity is known to be variable on decade-long time
scales, and \citet{2006AJ....132..161G} found $\log{\Rhk} = -4.85$ for
\eInd\ roughly eight years later. 

The available age calibrations of \Rhk\ give consistent estimates.
(1)~Using the \citet{1998csss...10.1235D} calibration, we find ages of
1.2~Gyr and 3.2~Gyr from the \citet{1996AJ....111..439H} and
\citet{2006AJ....132..161G} data, respectively. (2)~The
\citet{1991ApJ...375..722S} calibration results in ages of 1.2~Gyr and
3.4~Gyr.
(3)~Note that with a $B-V$ color of 1.06~mag
\citep{1997A&A...323L..49P}, \eInd~A is beyond the most recent
activity-age calibrations, those of \citet{mam08-ages} which are only
valid for $0.5 < B-V < 0.9$~mag stars. If we disregard this limitation,
their \Rhk\ calibration gives ages of 0.9 and 4.0~Gyr for the two sets
of measurements.
(4)~We can also use the ``preferred'' method of \citet{mam08-ages},
namely converting \Rhk\ to a Rossby number, in order to estimate a
rotation period that can then be used to derive an age from their
gyrochronology relation. The resulting ages are 1.6~Gyr and 5.0~Gyr for
the higher and lower activity levels, respectively.
The nominal uncertainty for such chromospheric ages is estimated to be
around $\pm50$\% at a single epoch \citep{1991ApJ...375..722S,
  mam08-ages}. Thus, the overall age range derived from chromospheric
indicators is 0.5--7.0~Gyr, with a mean of about 2~Gyr.

An alternative estimate comes from \citet{2009MNRAS.399..377J}. They
obtained high-contrast imaging of the system to attempt to directly
image a close-in substellar companion, whose existence has been inferred
from a long-term linear trend in the radial velocity of \eInd~A
\citep{2002A&A...392..671E}. By combining their imaging non-detection
combined with the radial velocity data, Janson \etal\ conclude that the
star is likely to be older than than $\approx$1~Gyr, or else the
companion would have been sufficiently self-luminous be have been
directly detected.

This age estimate may be improved in the future by measuring \eInd~A's
rotation period, as gyrochronology relations can provide more precise
age estimates \citep[15--25\%;][]{2007ApJ...669.1167B}.\footnote{Using
  the aforementioned \citet{1997MNRAS.284..803S}
  rotation period estimate,
  \citet{2007ApJ...669.1167B} compute a
  gyrochronological age of $1.03\pm0.13$~Gyr for \eInd~A. This age
  should be considered notional, since it relies on an estimated
  period, not a direct measurement.} Based on \citet{mam08-ages}, the
rotation period of \eInd~A is expected to be 14--50~days given its level 
of chromospheric activity.
Furthermore, because \eInd~A is a very bright star ($V = 4.7$~mag), it
is within reach of asteroseismological measurements with current
facilities, which may be able to provide an even more precise age
estimate.







\clearpage

\clearpage



\begin{deluxetable}{lccccccc}
\tablecaption{Keck LGS AO Observations \label{table:keck}}
\tabletypesize{\small}
\rotate
\tablewidth{0pt}
\tablehead{
  \colhead{Date} &
  \colhead{Filter\tablenotemark{a}} &
  \colhead{Airmass} &
  \colhead{FWHM} &
  \colhead{Strehl ratio} &
  \colhead{Separation} &
  \colhead{Position angle} &
  \colhead{$\Delta$mag} \\
  \colhead{(UT)} &
  \colhead{} &
  \colhead{} &
  \colhead{(mas)} &
  \colhead{} &
  \colhead{(mas)} &
  \colhead{(deg)} &
  \colhead{}
}

\startdata
2007-Apr-22  & $J$     & 1.46 &  112 $\pm$ 3  &  0.022 $\pm$ 0.002  & 143 $\pm$ 6\phn  &  311 $\pm$ 2  &  1.5 $\pm$ 0.2    \\
             & $H$     & 1.40 &  105 $\pm$ 5  &  0.049 $\pm$ 0.004  & 151 $\pm$ 13     &  314 $\pm$ 5  &  2.8 $\pm$ 0.3    \\
             & \Ks\    & 1.53 &\phn92 $\pm$ 5 &  0.121 $\pm$ 0.019  & 156 $\pm$ 7\phn  &  314 $\pm$ 4  &  3.2 $\pm$ 0.5    \\ 
2008-Jan-16  & $CH_4s$ & 1.22 &\phn77 $\pm$ 5 &  0.083 $\pm$ 0.012  & 142 $\pm$ 7\phn  &  313 $\pm$ 3  &  2.07 $\pm$ 0.17  \\
\enddata

\tablecomments{All photometry on the MKO system. The tabulated
  uncertainties on the imaging performance (FWHM and Strehl ratio)
  are the RMS of the measurements from the individual images.}

\end{deluxetable}

\clearpage
\begin{deluxetable}{lcc}
\tablecaption{Resolved Properties of \twomassbin{AB}\label{table:resolved}}
\tabletypesize{\small}
\tablewidth{0pt}
\tablehead{
\colhead{Property} &
\colhead{Component {A}} &
\colhead{Component {B}}
}

\startdata
$J$ (mag)                                                           &  15.80 $\pm$ 0.05     &  17.27 $\pm$ 0.17     \\
$CH_4s$ (mag)                                                       &  15.20 $\pm$ 0.04     &  17.27 $\pm$ 0.15     \\
$H$ (mag)                                                           &  15.32 $\pm$ 0.04     &  18.08 $\pm$ 0.26     \\
$K_S$ (mag)                                                         &  15.12 $\pm$ 0.14     &  18.29 $\pm$ 0.53     \\
$K$ (mag)                                                           &  15.22 $\pm$ 0.04     &  18.48 $\pm$ 0.52     \\
$J-H$ (mag)                                                         &  \phs0.48 $\pm$ 0.06  &   $-$0.81 $\pm$ 0.31  \\
$J-K$ (mag)                                                         &  \phs0.58 $\pm$ 0.07  &   $-$1.21 $\pm$ 0.54  \\
$CH_4s-H$ (mag)                                                     &   $-$0.12 $\pm$ 0.05  &   $-$0.81 $\pm$ 0.30  \\
$H-K$ (mag)                                                         &  \phs0.10 $\pm$ 0.05  &   $-$0.40 $\pm$ 0.58  \\
Estimated spectral type                                             &    T2.0 $\pm$ 0.5     &    T7.5 $\pm$ 0.5     \\
$d_{phot}$ (pc)                                                      &     \multicolumn{2}{c}{21 $\pm$  4}           \\
$M(J)$ (mag)                                                        &  14.19 $\pm$ 0.41     &  15.66 $\pm$ 0.44     \\
$M(CH_4s)$ (mag)                                                    &  13.59 $\pm$ 0.41     &  15.66 $\pm$ 0.44     \\
$M(H)$ (mag)                                                        &  13.71 $\pm$ 0.41     &  16.47 $\pm$ 0.49     \\
$M(K_S)$ (mag)                                                      &  13.51 $\pm$ 0.43     &  16.68 $\pm$ 0.67     \\
$M(K)$ (mag)                                                        &  13.61 $\pm$ 0.41     &  16.87 $\pm$ 0.66     \\
$\log(L_{bol,A+B}/\Lsun)$                                            &  \multicolumn{2}{c}{$-4.61\pm0.15(0.02)$\tablenotemark{a}} \\
$M_{bol}$  (from $J$-band)                                           &  16.21 $\pm$ 0.45     &  18.24 $\pm$ 0.47     \\
$\log(L_{bol}/L_{\odot})$ (from $J$-band)                             &  $-$4.58 $\pm$ 0.18   &  $-$5.39 $\pm$ 0.19   \\
$\log(L_{bol,A} / L_{bol,B})$ (from $J$-band)                         &  \multicolumn{2}{c}{0.81 $\pm$ 0.11}            \\

\enddata

\tablecomments{All infrared photometry on the MKO photometric system.
  We also measure integrated-light \Spitzer/IRAC
    photometry for the system of $[3.6]=14.02\pm0.03$,
    $[4.5]=13.49\pm0.03$, $[5.8]=13.33\pm0.03$, and
    $[8.0]=13.06\pm0.07$~mags (\S~\ref{sec:lbol}).}

\tablenotetext{a}{The error in parenthesis gives the \Lbol\ error
  assuming there is no uncertainty in the distance.}

\end{deluxetable}


\clearpage
\begin{deluxetable}{lcc}
  \tablecaption{Properties of \twomassbin{AB} Derived from Evolutionary
    Models \label{table:evolmodels}}
\tabletypesize{\small}
\tablewidth{0pt}
\tablehead{
\colhead{Property} &
\colhead{Component {A}} &
\colhead{Component {B}}
}

\startdata

\cutinhead{$t = 0.5$~Gyr} 
Total mass ($M_{Jup}$)                                               &    \multicolumn{2}{c}{52 $\pm$ 7}              \\
Mass ($M_{Jup}$)                                                     &     36 $\pm$    6     &     17 $\pm$    4     \\
$q\ (\equiv M_B/M_A)$                                               &   \multicolumn{2}{c}{0.47 $\pm$ 0.16}         \\
$\Teff (K)$                                                        &   1281 $\pm$  162     &    771 $\pm$  103     \\
$\Teff_{,B}/\Teff_{,A}$                                               &  \multicolumn{2}{c}{0.60 $\pm$ 0.08}          \\
Radius $(R_\odot)$                                                   &  0.101 $\pm$ 0.002    &  0.109 $\pm$ 0.003    \\
$\log(g)$ (cgs)                                                      &   4.96 $\pm$ 0.11     &   4.56 $\pm$ 0.13     \\
Orbital period (yr)                                                  &   \multicolumn{2}{c}{27$^{+50}_{-10}$}          \\

\cutinhead{$t = 1$~Gyr}  
Total mass ($M_{Jup}$)                                               &   \multicolumn{2}{c}{75 $\pm$ 10}             \\
Mass ($M_{Jup}$)                                                     &     49 $\pm$    9     &     25 $\pm$    5     \\
$q\ (\equiv M_B/M_A)$                                               &   \multicolumn{2}{c}{0.51 $\pm$ 0.10}          \\
$T_{eff} (K)$                                                        &   1336 $\pm$  173     &    802 $\pm$  107     \\
$\Teff_{,B}/\Teff_{,A}$                                              &   \multicolumn{2}{c}{0.60 $\pm$ 0.07}          \\
Radius $(R_\odot)$                                                  &  0.092 $\pm$ 0.002    &  0.101 $\pm$ 0.003    \\
$\log(g)$ (cgs)                                                     &   5.18 $\pm$ 0.11     &   4.80 $\pm$ 0.12     \\
Orbital period (yr)                                                 &  \multicolumn{2}{c}{22$^{+42}_{-8}$}           \\    

\cutinhead{$t = 5$~Gyr}   
Total mass ($M_{Jup}$)                                               &   \multicolumn{2}{c}{132 $\pm$ 10}               \\
Mass ($M_{Jup}$)                                                     &     77 $\pm$    5     &     55 $\pm$    8     \\
$q\ (\equiv M_B/M_A)$                                                &    \multicolumn{2}{c}{0.72 $\pm$ 0.07}           \\
$T_{eff} (K)$                                                        &   1418 $\pm$  163     &    881 $\pm$  120     \\
$\Teff_B/\Teff_A$                                                    &   \multicolumn{2}{c}{0.62 $\pm$ 0.07}            \\
Radius $(R_\odot)$                                                   &  0.082 $\pm$ 0.001    &  0.083 $\pm$ 0.003    \\
$\log(g)$ (cgs)                                                     &   5.48 $\pm$ 0.04     &   5.31 $\pm$ 0.11     \\
Orbital period (yr)                                                 &    \multicolumn{2}{c}{17$^{+31}_{-6}$}          \\

\cutinhead{$t = 10$~Gyr}   
Total mass ($M_{Jup}$)                                              &    \multicolumn{2}{c}{149 $\pm$    8}          \\
Mass ($M_{Jup}$)                                                     &     79 $\pm$    3       &     70 $\pm$    7     \\
$q\ (\equiv M_B/M_A)$                                               &   \multicolumn{2}{c}{0.90 $\pm$ 0.03}         \\
$T_{eff} (K)$                                                        &   1424 $\pm$  159       &    908 $\pm$  119     \\
$\Teff_{,B}/\Teff_{,A}$                                               &   \multicolumn{2}{c}{0.64 $\pm$ 0.07}          \\
Radius $(R_\odot)$                                                    &  0.081 $\pm$ 0.002      &  0.078 $\pm$ 0.002    \\
$\log(g)$ (cgs)                                                      &   5.50 $\pm$ 0.03       &   5.47 $\pm$ 0.09     \\
Orbital period (yr)                                                  &   \multicolumn{2}{c}{15.9$^{+29.4}_{-5.8}$}      \\

\enddata

\end{deluxetable}

\clearpage

\thispagestyle{empty}
\begin{deluxetable}{lccccclllccccccc}
\tablecaption{Homogenous Compilation of Substellar Field Binaries \label{table:binaries}}
\tabletypesize{\tiny}
\rotate
\tablewidth{0pt}
\setlength{\tabcolsep}{0.015in} 
\tablehead{
  \colhead{Binary} &
  \colhead{SpT(A+B)}   &
  \colhead{SpT(A)}   &
  \colhead{SpT(B)}   &
  \colhead{$d$ (pc)}   &
  \colhead{Age (Gyr)} &
  \colhead{$\log(\Lbol_{,A})$}   &
  \colhead{$\log(\Lbol_{,B})$}   &
  \colhead{$\log(\frac{\Lbol_{,A}}{\Lbol_{,B}})$} &  
  \colhead{$q$} &
  \multicolumn{5}{c}{References} &
  \colhead{Notes} \\
\cline{11-15} 
  \colhead{} &
 \colhead{}   &
  \colhead{}   &
  \colhead{}   &
  \colhead{}   &
  \colhead{} &
  \colhead{}   &
  \colhead{}   &
  \colhead{}  &
  \colhead{}  &
  \colhead{Disc} &
  \colhead{Phot} &
  \colhead{SpT}   &
  \colhead{Dist}   &
  \colhead{Age}   &
 \colhead{}
}

\startdata

DENIS-P~J004135.3$-$562112AB   & M$8.0\pm0.5$ & M$6.5\pm1.0$ & M$9.0\pm1.0$         & $50_{-10}^{+21}$             & 0.01 (0.005--0.2)         & $-2.44\pm0.34(0.03)$  &  $-2.78\pm0.34(0.04)$   &  $0.34\pm0.05$  &   0.74$\pm$0.18  & 55     & 16, 55        & 53, 55     & 55     & 54     &      \\ 
LP~349-25AB                    & M$8.0\pm0.5$ & M$7.5\pm1.0$ & M$8.0^{+1.0}_{-1.0}$ & $13.2\pm0.3$                 & $0.127_{-0.017}^{+0.021}$ & $-3.04\pm0.02(0.02)$  &  $-3.18\pm0.03(0.03)$   &  $0.13\pm0.02$  &  0.872$_{-0.018}^{+0.014}$  & 21     & 16, 19        & 19, 23     & 22     & 19     & a  \\ 
Gl~569Bab                      & M$8.5\pm0.5$ & M$8.5\pm0.5$ & M$9.0\pm0.5$         & $9.65\pm0.16$                & $0.46_{-0.07}^{+0.11}$    & $-3.42\pm0.02(0.02)$  &  $-3.62\pm0.02(0.02)$   &  $0.20\pm0.02$  &  0.866$_{-0.014}^{+0.019}$  & 46     & 19, 36        & 28, 36     & 61     & 19     & a  \\  
DENIS-P~J035726.9$-$441730AB   & L$0.0\pm0.5$ & M$9.0\pm0.5$ & L$1.5\pm0.5$         & $40\pm7$                     & $\sim$0.1                 & $-3.58\pm0.15(0.04)$  &  $-3.77\pm0.15(0.04)$   &  $0.19\pm0.05$  &   0.87$\pm$0.03  & 3      & 4, 16         & 15, 47     &        & 15     & b  \\ 
Kelu-1AB                       & L$2.0\pm0.5$ & L$2.0\pm0.5$ & L$3.5\pm0.5$         & $18.7\pm0.7$                 & 0.5 (0.3--0.8)            & $-3.82\pm0.05(0.04)$  &  $-4.00\pm0.05(0.04)$   &  $0.18\pm0.04$  &   0.92$\pm$0.02  & 39     & 37, 39        & 31         & 17     & 39     &      \\ 
2MASS~J11463449$+$2230527AB    & L$3.0\pm0.5$ & L$3.0\pm0.5$ & L$3.0\pm1.0$         & $27.2\pm0.6$                 & 0.5 (0.3--0.8)            & $-3.82\pm0.04(0.04)$  &  $-3.96\pm0.04(0.04)$   &  $0.13\pm0.05$  &   0.94$\pm$0.02  & 50     & 4, 16         & 31         & 17     & 1      & b,c \\ 
SDSS~J224953.47$+$004404.6AB   & L$3.0\pm1.0$ & L$3.0\pm0.5$ & L$5.0\pm1.0$         & $53\pm15$                    & 0.1 (0.02--0.3)           & $-3.9 \pm0.3 (0.03)$  &  $-4.2 \pm0.3 (0.04)$   &  $0.36\pm0.02$  &   0.75$\pm$0.10  & 2      & 2, 34         & 2, 27      & 2      & 2      &      \\  
2MASS~J00250365$+$4759191AB    & L$4.0\pm1.0$ & L$4.0\pm1.0$ & L$4.5_{-1.0}^{+2.0}$ & $25\pm7$                     & 0.3 (0.1--0.5)            & $-4.08\pm0.20(0.04)$  &  $-4.13\pm0.22(0.04)$   &  $0.06\pm0.06$  &   0.96$\pm$0.04  & 51     & 9, 16, 51     & 14         &        & 20     & d  \\ 
HD~130948BC                    & L$4.0\pm1.0$ & L$4.0\pm1.0$ & L$4.0\pm1.0$         & $18.17\pm0.11$               & $0.79_{-0.23}^{+0.22}$    & $-3.82\pm0.05(0.04)$  &  $-3.90\pm0.05(0.04)$   &  $0.08\pm0.01$  &  0.948$\pm$0.005 & 49     & 18            & 26         & 61     & 18     & a  \\  
Gl~417BC                       & L$4.5\pm0.5$ & L$4.5\pm0.5$ & L$4.5_{-0.5}^{+1.0}$ & $21.9\pm0.2$                 & 0.74 (0.58--0.86)         & $-4.06\pm0.03(0.03)$  &  $-4.18\pm0.04(0.04)$   &  $0.12\pm0.05$  &   0.94$\pm$0.02  & 3      & 16, 42        & 32         & 61     & 2      &      \\ 
GJ~1001BC                      & L$5.0\pm0.5$ & L$5.0\pm0.5$ & L$5.0\pm1.0$         & $13.0_{-0.6}^{+0.7}$         & 3 (1--10)                 & $-4.03\pm0.06(0.05)$  &  $-4.08\pm0.07(0.05)$   &  $0.05\pm0.08$  &   0.99$\pm$0.01  & 25     & 16, 25        & 33         & 29     & 38     & e  \\ 
DENIS-P~J1228.2$-$1547AB       & L$5.0\pm0.5$ & L$5.0\pm0.5$ & L$5.0\pm1.0$         & $20.2_{-0.7}^{+0.8}$         &  			       & $-4.19\pm0.05(0.03)$  &  $-4.24\pm0.05(0.04)$   &  $0.05\pm0.05$  &   0.98$\pm$0.03  & 35, 45 & 16, 42        & 31         & 17     &        &      \\ 
2MASS~J12392727$+$5515371AB    & L$5.0\pm0.5$ & L$5.0\pm0.5$ & L$6.0_{-1.5}^{+2.5}$ & $19\pm4$                     &  			       & $-4.23\pm0.17(0.04)$  &  $-4.31\pm0.17(0.06)$   &  $0.09\pm0.07$  &   0.95$\pm$0.05  & 24     & 3, 16         & 32         &        &        & b  \\ 
2MASS~J21321145$+$1341584AB    & L$6.0\pm0.5$ & L$5.0\pm0.5$ & L$7.5\pm0.5$         & $23\pm4$                     &  			       & $-4.39\pm0.17(0.04)$  &  $-4.72\pm0.17(0.04)$   &  $0.33\pm0.04$  &   0.76$\pm$0.03  & 58     & 16, 58        & 14, 58     &        &        &      \\ 
2MASS~J17281150$+$3948593AB    & L$7.0\pm0.5$ & L$7.0\pm0.5$ & L$9.0_{-1.0}^{+2.0}$ & $24\pm2$                     &  			       & $-4.47\pm0.08(0.04)$  &  $-4.65\pm0.08(0.04)$   &  $0.18\pm0.04$  &   0.85$\pm$0.03  & 24     & 16, 42        & 32         & 62     &        &      \\ 
2MASS~J08503593$+$1057156AB    & L$6.0\pm0.5$ & L$6.0\pm0.5$ & L$9.5\pm1.0$         & $26_{-2}^{+3}$               &  			       & $-4.53\pm0.08(0.03)$  &  $-4.87\pm0.09(0.03)$   &  $0.34\pm0.04$  &   0.75$\pm$0.03  & 50     & 37, 42        & 31         & 17     &        &      \\ 
2MASS~J22551861$-$5713056AB    & L$6.0\pm1.0$ & L$5.0\pm1.0$ & L$8.0\pm1.0$         & $12\pm3$                     &  			       & $-4.38\pm0.20(0.04)$  &  $-4.95\pm0.22(0.07)$   &  $0.57\pm0.09$  &   0.63$\pm$0.05  & 53     & 9, 16, 51     & 53         &        &        &      \\ 
2MASS~J21522609$+$0937575AB    & L$6.0\pm1.0$ & L$6.0\pm1.0$ & L$8.5_{-3.0}^{+4.0}$ & $22\pm5$                     &  			       & $-4.36\pm0.19(0.04)$  &  $-4.39\pm0.20(0.04)$   &  $0.03\pm0.05$  &   0.98$\pm$0.04  & 51     & 16, 51        & 53         &        &        &      \\ 
SDSS~J141624.08$+$134826.7AB   & L$6.0\pm0.5$ & L$6.0\pm0.5$ & T$7.5\pm0.5$         & $8_{-1}^{+2}$                &  			       & $-4.30\pm0.20(0.03)$  &  $-6.24\pm0.20(0.03)$   &  $1.94\pm0.04$  &   0.18$\pm$0.01  & 12     & 12            & 5, 12      & 57     &        &  f  \\ 
DENIS-P~J225210.7$-$173013AB   & L$7.5\pm1.0$ & L$4.5\pm1.0$ & T$4.5\pm1.0$         & $11\pm2$                     &  			       & $-4.51\pm0.18(0.04)$  &  $-4.93\pm0.18(0.04)$   &  $0.42\pm0.05$  &   0.70$\pm$0.03  & 52     & 9, 16, 42, 51 & 11, 30     &        &        &  g \\ 
2MASS~J05185995$-$2828372AB    & T$1.0\pm0.5$ & L$7.5\pm0.5$ & T$5.0\pm0.5$         & $20\pm4$                     &  			       & $-4.55\pm0.19(0.10)$  &  $-5.29\pm0.23(0.15)$   &  $0.75\pm0.22$  &   0.54$\pm$0.10  & 9      & 9, 16         & 10, 11     &        &        &      \\ 
SDSS~J042348.57$-$041403.5AB   & T$0.0\pm0.5$ & L$7.5\pm0.5$ & T$2.0\pm0.5$         & $15.2\pm0.4$                 &  			       & $-4.33\pm0.04(0.03)$  &  $-4.59\pm0.04(0.03)$   &  $0.26\pm0.04$  &   0.80$\pm$0.03  & 7      & 9, 37         & 10, 11     & 62     &        &      \\ 
DENIS-P~J020529.0$-$115925AB   & L$7.0\pm0.5$ & L$7.0\pm0.5$ & L$6.0\pm1.0$         & $19.8\pm0.6$                 &  			       & $-4.28\pm0.04(0.04)$  &  $-4.34\pm0.05(0.04)$   &  $0.06\pm0.05$  &   0.96$\pm$0.03  & 35     & 16, 42        & 31         & 17     &        &      \\ 
2MASS~J09153413$+$0422045AB    & L$6.0\pm1.0$ & L$6.0\pm1.0$ & L$7.0_{-1.5}^{+3.0}$ & $18\pm5$                     &  			       & $-4.28\pm0.20(0.04)$  &  $-4.33\pm0.20(0.04)$   &  $0.05\pm0.05$  &   0.97$\pm$0.04  & 51     & 9, 16, 42, 51 & 53         &        &        &      \\ 
2MASS~J21011544$+$1756586AB    & L$7.5\pm0.5$ & L$7.5\pm0.5$ & T$1.0_{-2.5}^{+2.0}$ & $33_{-3}^{+4}$               &  			       & $-4.55\pm0.11(0.04)$  &  $-4.67\pm0.13(0.08)$   &  $0.13\pm0.09$  &   0.89$\pm$0.07  & 3      & 3, 13         & 32         & 62     &        & b  \\ 
2MASS~J03105986$+$1648155AB    & L$8.0\pm0.5$ & L$8.0\pm0.5$ & L$9.0_{-1.5}^{+2.5}$ & $25\pm4$                     &  			       & $-4.54\pm0.15(0.03)$  &  $-4.60\pm0.15(0.04)$   &  $0.06\pm0.04$  &   0.95$\pm$0.03  & 59     & 37, 42        & 32         &        &        &      \\ 
Gl~337CD                       & T$0.0\pm0.5$ & T$0.0\pm0.5$ & T$0.5_{-2.0}^{+1.5}$ & $20.4\pm0.2$                 & 1.5 (0.6--3.4)            & $-4.62\pm0.04(0.04)$  &  $-4.70\pm0.05(0.05)$   &  $0.08\pm0.04$  &   0.93$\pm$0.03  & 8      & 16, 42        & 10, 63     & 61     & 63     &      \\ 
2MASS~J09201223$+$3517429AB    & T$0.0\pm0.5$ & T$0.0\pm0.5$ & T$1.0_{-2.0}^{+2.5}$ & $18\pm3$                     &  			       & $-4.75\pm0.15(0.04)$  &  $-4.84\pm0.15(0.04)$   &  $0.09\pm0.05$  &   0.93$\pm$0.04  & 50     & 16, 42        & 10, 32     &        &        &  h \\ 
SDSS~J105213.51$+$442255.7AB   & T$0.5\pm1.0$ & T$0.5\pm1.0$ & T$0.5_{-2.0}^{+2.5}$ & $22\pm4$                     &  			       & $-4.76\pm0.16(0.03)$  &  $-4.78\pm0.16(0.04)$   &  $0.02\pm0.05$  &   0.98$\pm$0.04  & 42     & 13, 42        & 13         &        &        &      \\ 
SDSS~J205235.31$-$160929.8AB   & T$1.0\pm1.0$ & T$1.0\pm1.0$ & T$3.5_{-2.0}^{+1.5}$ & $25\pm5$                     &  			       & $-4.74\pm0.16(0.03)$  &  $-4.85\pm0.16(0.04)$   &  $0.12\pm0.05$  &   0.90$\pm$0.04  & 42     & 13, 42        & 13         &        &        &      \\ 
2MASS~J14044948$-$3159330AB    & T$2.5\pm0.5$ & T$0.0\pm0.5$ & T$5.0\pm0.5$         & $18\pm3$                     &  			       & $-4.84\pm0.16(0.05)$  &  $-4.99\pm0.16(0.05)$   &  $0.16\pm0.06$  &   0.88$\pm$0.04  & 44     & 44            & 11, 43     &        &        &      \\ 
SDSS~J102109.69$-$030420.1AB   & T$3.0\pm0.5$ & T$1.0\pm0.5$ & T$5.5\pm0.5$         & $29_{-3}^{+5}$               &  			       & $-4.58\pm0.13(0.04)$  &  $-4.75\pm0.13(0.05)$   &  $0.17\pm0.06$  &   0.86$\pm$0.05  & 9      & 9, 37, 42     & 10, 11     & 60     &        &      \\ 
$\epsilon$~Ind~Bab             & T$2.5\pm0.5$ & T$1.0\pm0.5$ & T$6.0\pm0.5$         & $3.622\pm0.004$              & 2 (0.5--7.0)              & $-4.62\pm0.03(0.03)$  &  $-5.24\pm0.04(0.04)$   &  $0.61\pm0.05$  &   0.63$\pm$0.02  & 48     & 16, 48        & 10, 48, 56 & 61     & 1      &      \\ 
SDSS~J153417.05$+$161546.1AB   & T$3.5\pm0.5$ & T$1.5\pm0.5$ & T$5.0\pm0.5$         & $31\pm6$                     &  			       & $-4.87\pm0.16(0.03)$  &  $-5.09\pm0.16(0.04)$   &  $0.23\pm0.05$  &   0.83$\pm$0.03  & 40     & 13, 40        & 40         &        &        &      \\ 
2MASS~J12095613$-$1004008AB    & T$3.0\pm0.5$ & T$2.0\pm0.5$ & T$7.5\pm0.5$         & $21\pm4$                     &  			       & $-4.58\pm0.18(0.07)$  &  $-5.39\pm0.19(0.09)$   &  $0.81\pm0.11$  &   0.51$\pm$0.05  & 1      & 1, 13         & 1, 10      &        &        &      \\ 
SDSS~J092615.38$+$584720.9AB   & T$4.5\pm0.5$ & T$4.5\pm0.5$ & T$3.0_{-4.0}^{+2.5}$ & $22\pm4$                     &  			       & $-4.79\pm0.16(0.06)$  &  $-4.95\pm0.21(0.08)$   &  $0.16\pm0.13$  &   0.88$\pm$0.09  & 9      & 9, 37         & 10         &        &        &      \\ 
2MASS~J15344984$-$2952274AB    & T$5.0\pm0.5$ & T$5.0\pm0.5$ & T$5.5\pm0.5$         & $13.6\pm0.2$                 & $0.78_{-0.09}^{+0.09}$    & $-5.02\pm0.02(0.02)$  &  $-5.09\pm0.02(0.02)$   &  $0.08\pm0.02$  &  0.936$\pm$0.012 & 6      & 34, 41        & 10, 41     & 60     & 41     &  a \\ 
2MASS~J12255432$-$2739476AB    & T$6.0\pm0.5$ & T$6.0\pm0.5$ & T$7.5\pm0.5$         & $13.3_{-0.4}^{+0.5}$         &  			       & $-4.93\pm0.05(0.05)$  &  $-5.47\pm0.05(0.04)$   &  $0.54\pm0.06$  &   0.63$\pm$0.03  & 6      & 37, 42        & 10         & 60     &        &      \\ 
2MASS~J15530228$+$1532369AB    & T$7.0\pm0.5$ & T$7.0\pm0.5$ & T$7.5\pm0.5$         & $11\pm3$                     &  			       & $-5.42\pm0.24(0.04)$  &  $-5.54\pm0.25(0.03)$   &  $0.12\pm0.05$  &   0.90$\pm$0.05  & 9      & 9, 34, 42     & 10         &        &        &      \\

\enddata

\tablecomments{Field binaries are included in this table if the primary
  spectral type is L4 or later, or if there is other evidence that the
  system is young enough to be substellar, \ie, from accretion
  signatures (DENIS-P~J0041$-$56AB), lithium absorption (Kelu-1 and
  2MASS~J1146$+$22AB), low surface gravity (DENIS-P~J0357$-$44AB and
  SDSS~J2249$+$00AB), or dynamical masses (LP~349-25AB and Gl~569Bab).
  Distance references are listed for objects with trigonometric
  parallaxes or those with a wide main-sequence companion with a good
  distance estimate; otherwise the distances are photometric estimates
  derived by us, except DENIS-P~J0041$-$56AB comes from Reiners (2009).
  The uncertainties on \Lbol\ are listed with and without the
  uncertainty in the distance folded in. Other notes: (a)~The mass
  ratios come from the published analysis using the measured total
  dynamical mass and bolometric luminosity of the system combined with
  the luminosity ratio of the two components. (b)~The $K$-band flux
  ratio was estimated from $F814W$-band flux ratio (Appendix~B). (c)~The
  age estimate assigned by us is based on the presence of lithium and
  the \Lbol\ of the components using evolutionary models (\eg,
  \citealp{2005astro.ph..8082L}). (d)~This object also has a distance
  estimate from a comoving wide F8 companion star (G~171-58) of
  42.2$^{+2.0}_{-1.8}$~pc (Faherty \etal\ 2010), which is somewhat
  discrepant with our photometric distance listed here. (e)~GJ~1001BC
  has an integrated light spectral type of L5.0$\pm$0.5 and essentially
  identical fluxes and colors; thus both components are likely to be L5
  \citep{2004AJ....128.1733G}. The absolute magnitudes of component~C
  formally given to a spectral type of L4.0$\pm$1.0 using our
  methodology in Appendix~\ref{sec:binaries}, which is consistent. We
  adopt L5.0$\pm$0.5 for both components. (f)
  \citet{2010MNRAS.404.1952B} and \citet{2010AJ....139.2448B} find the
  secondary is high gravity and low temperature and suggest an age of
  $\sim$10~Gyr and 2--10~Gyr, respectively. For an age of 10~Gyr, the
  mass ratio of the binary would be $\approx$0.5, based on their mass
  estimates for component~B and that of \citet{2010ApJ...710...45B} for
  component~A. (g) We list the IR spectral type, as an optical type is
  not available for this object. (h) This has very disparate
  integrated-light spectral types, L6.5 in the optical and T0 in the
  near-IR. We use the IR type here.}

\tablerefs{
 (1)   This work;		       
 (2)   \citealp{allers09-sdss2249};    
 (3)   \citealp{2003AJ....126.1526B};  
 (4)   \citealp{2008A&A...481..757B};  
 (5)   \citealp{2010ApJ...710...45B};  
 (6)   \citealp{2003ApJ...586..512B};  
 (7)   \citealp{2005ApJ...634L.177B};  
 (8)   \citealp{2005AJ....129.2849B};  
 (9)   \citealp{2006ApJS..166..585B};  
(10)   \citealp{2006ApJ...637.1067B};  
(11)   \citealp{2010ApJ...710.1142B};  
(12)   \citealp{2010MNRAS.404.1952B};  
(13)   \citealp{chiu05};	       
(14)   \citealp{2007AJ....133..439C};  
(15)   \citealp{2009AJ....137.3345C};  
(16)   \citealp{2003tmc..book.....C};  
(17)   \citealp{2002AJ....124.1170D};  
(18)   \citealp{2008arXiv0807.2450D};  
(19)   \citealp{2010arXiv1007.4197D};  
(20)   \citealp{2010AJ....139..176F};  
(21)   \citealp{2005A&A...435L...5F};  
(22)   \citealp{2009AJ....137..402G};  
(23)   \citealp{2000AJ....120.1085G};  
(24)   \citealp{2003AJ....125.3302G};  
(25)   \citealp{2004AJ....128.1733G};  
(26)   \citealp{2002ApJ...567L..59G};  
(27)   \citealp{2002astro.ph..4065H};  
(28)   \citealp{1990ApJ...354L..29H};  
(29)   \citealp{2006AJ....132.2360H};  
(30)   \citealp{2004A&A...416L..17K};  
(31)   \citealp{1999ApJ...519..802K};  
(32)   \citealp{2000AJ....120..447K};  
(33)   \citealp{2001AJ....121.3235K};  
(34)   \citealp{2004AJ....127.3553K};  
(35)   \citealp{1999ApJ...526L..25K};  
(36)   \citealp{2001ApJ...560..390L};  
(37)   \citealp{leg01}; 	       
(38)   \citealp{2002MNRAS.332...78L};  
(39)   \citealp{2005astro.ph..8082L};  
(40)   \citealp{2006astro.ph..5037L};  
(41)   \citealp{liu08-2m1534orbit};    
(42)   Liu et al., in prep;            
(43)   \citealp{2007AJ....134.1162L};  
(44)   \citealp{2008ApJ...685.1183L};  
(45)   \citealp{1999Sci...283.1718M};  
(46)   \citealp{2000ApJ...529L..37M};  
(47)   \citealp{2006A&A...456..253M};  
(48)   \citealp{2004A&A...413.1029M};  
(49)   \citealp{2002ApJ...567L.133P};  
(50)   \citealp{2001AJ....121..489R};  
(51)   \citealp{2006AJ....132..891R};  
(52)   \citealp{2006ApJ...639.1114R};  
(53)   \citealp{2008AJ....136.1290R};  
(54)   \citealp{2009ApJ...702L.119R};  
(55)   \citealp{2010A&A...513L...9R};  
(56)   \citealp{2003A&A...398L..29S};  
(57)   \citealp{2010A&A...510L...8S};  
(58)   \citealp{2007AJ....133.2320S};  
(59)   \citealp{2010A&A...516A..37S};  
(60)   \citealp{2003AJ....126..975T};  
(61)   \citealp{2007A&A...474..653V};  
(62)   \citealp{2004AJ....127.2948V};  
(63)   \citealp{wils01}.	       
}

\end{deluxetable}

\clearpage
\thispagestyle{empty}
\begin{deluxetable}{llccccccc}
\tablecaption{Ages and Masses Inferred from the H-R Diagram \label{table:ages}}
\tabletypesize{\small}
\rotate
\setlength{\tabcolsep}{0.035in} 
\tablewidth{0pt}
\tablehead{
  \colhead{$T_{eff,A}^{atm},\ T_{eff,B}^{atm}$} &
  \colhead{Model} &
  \colhead{$\log (t_A/{\rm yr})$} &
  \colhead{$\log (t_B/{\rm yr})$} &
  \colhead{$\log (t_A/t_B)$} &
  \colhead{$M_A$} &
  \colhead{$M_B$} &
  \colhead{$M_B + M_A$} &
  \colhead{$q \equiv M_B/M_A$} \\
  \colhead{} &
  \colhead{} &
  \colhead{} &
  \colhead{} &
  \colhead{} &
  \colhead{(\Mjup)} &
  \colhead{(\Mjup)} &
  \colhead{(\Mjup)} &
  \colhead{} 
}

\startdata

\multicolumn{9}{c}{\bf \twomassbin{AB}} \\
\hline

$1150\pm100$~K, $800\pm70$~K (this work)  &  Lyon/COND   &  7.5$_{-1.5}^{+1.2}$  &  8.5$_{-1.5}^{+1.0}$  &  $-$0.8$_{-1.3}^{+1.3}$ 
                                                         &  10$_{-7}^{+16}$     &  18$_{-13}^{+25}$     &  34$_{-19}^{+27}$      &  1.7$_{-1.3}^{+5.2}$  \\
                              &  Tucson      &  7.6$_{-1.5}^{+1.2}$  &  8.8$_{-1.7}^{+1.1}$  &  $-$1.0$_{-1.3}^{+1.2}$ 
                                             &  11$_{-8}^{+24}$     &  20$_{-17}^{+43}$     &  36$_{-25}^{+63}$      &  1.6$_{-1.3}^{+6.2}$  \\
\hline
& & & & & & & & \\
\multicolumn{9}{c}{\bf {\eInd~Bab}} \\
\hline
$1275\pm25$~K, $900\pm25$~K (Kasper \etal\ 2009)  &  Lyon/COND  &  9.1$\pm$0.2    &  8.8$\pm$0.3          &  0.2$\pm$0.3  
                                                               &  44$_{-6}^{+8}$  &  27$_{-7\phn}^{+7\phn}$  &  72$_{-9}^{+10}$  &  0.6$\pm$0.2  \\
                             &  Tucson     &  9.1$\pm$0.2  &  9.0$\pm$0.2    &  0.1$\pm$0.3
                                           &  49$\pm$8     &  28$_{-7}^{+8}$  &  78$_{-10}^{+11}$  &  0.6$\pm$0.2  \\

$1320\pm20$~K, $910\pm30$~K (King \etal\ 2010)  &  Lyon/COND  &  9.4$\pm$0.2  &  8.9$\pm$0.3   &  0.5$_{-0.3}^{+0.4}$
                                                              &  57$\pm$7     &  30$\pm$8      &  86$_{-10}^{+11}$  &  0.52$_{-0.14}^{+0.17}$  \\
                             &  Tucson     &  9.3$\pm$0.2    &  9.0$\pm$0.3     &  0.3$_{-0.4}^{+0.3}$
                                           &  63$_{-7}^{+8}$  &  31$_{-9}^{+11}$  &  95$_{-12}^{+13}$  &  0.49$_{-0.14}^{+0.19}$  \\

\enddata

\tablecomments{The results for each component are the median and
  68\% confidence limits from the Monte Carlo-generated probability
  distributions for the ages and masses computed from the H-R diagram
  positions (\S~\ref{sec:isochrone}).}

\end{deluxetable}

\clearpage
\thispagestyle{empty}
\begin{deluxetable}{lccccccccc}
\tablecaption{Coefficients of Polynomial Fits for Bolometric Correction Versus
  Near-IR Spectral Type \label{table:bolcorr}}
\rotate
\tabletypesize{\small}
\setlength{\tabcolsep}{0.04in} 
\tablewidth{0pt}
\tablehead{
  \colhead{Bandpass} &
  \colhead{$c_0$} &
  \colhead{$c_1$} &
  \colhead{$c_2$} &
  \colhead{$c_3$} &
  \colhead{$c_4$} &
  \colhead{$c_5$} &
  \colhead{$c_6$} &
  \colhead{RMS (mag)} 
}

\startdata

$J_{MKO}$ & 1.890448e+01 & $-$8.053993e+00 & 1.491738e+00   & $-$1.367605e$-$01 & 6.540717e$-$03 & $-$1.558986e$-$04 & 1.462266e$-$06 & 0.14 \\
$H_{MKO}$ & 1.366709e+01 & $-$5.426683e+00 & 1.027185e+00   & $-$9.618535e$-$02 & 4.733874e$-$03 & $-$1.171595e$-$04 & 1.148133e$-$06 & 0.07 \\
$K_{MKO}$ & 5.795845e+00 & $-$1.471358e+00 & 2.868014e$-$01 & $-$2.647474e$-$02 & 1.282657e$-$03 & $-$3.177629e$-$05 & 3.159780e$-$07 & 0.08 \\

\enddata

\tablecomments{These are the coefficients of the 6th-order polynomial
  fits plotted in Figure~\ref{fig:bolcorr} for the bolometric
  correction as a function of near-IR spectral type.  The fit for
  filter $X$ is defined as $$BC(X) = \sum_{i=0}^6 c_i \times
  (SpT)^{i}$$ where the numerical spectral type is defined as $SpT=11$
  for L1, $SpT=12$ for L2, $SpT=20$ for T0, etc.  The fits are
  applicable from M6 to T8.5.  The L~dwarfs are classified on the
  \citet{geb01} scheme and the T~dwarfs on the
  \citet{2005astro.ph.10090B} scheme.  The last column gives the
  standard deviation about the fit in magnitudes.}

\end{deluxetable}


\end{document}